\documentclass[usenatbib, twocolumn, nofootinbib]{aastex631}

\usepackage{multirow}
\usepackage{natbib}
\usepackage{color}
\usepackage{amsmath,amsfonts,bm}
\usepackage{hyperref}
\usepackage[T1]{fontenc}
\usepackage{graphicx}	
\usepackage[mathlines]{lineno}
\usepackage{wrapfig}
\usepackage{fontawesome}

\newcolumntype{C}[1]{>{\centering\let\newline\\\arraybackslash\hspace{0pt}}m{#1}}

\newcommand{\cmbsfour}{CMB-S4}
\newcommand{\sfour}{S4-Wide}
\newcommand{\sfourdeep}{S4-Ultra deep}
\newcommand{\cmbhd}{\mbox{CMB-HD}}
\newcommand{\spt}{SPT}
\newcommand{\sptsz}{SPT-SZ}
\newcommand{\sptpol}{SPTpol}
\newcommand{\sptthreeg}{SPT-3G}

\newcommand{\planck}{{\it Planck}}

\newcommand{\snr}{S/N}
\newcommand{\simonsobs}{SO}
\newcommand{\sogoal}{SO-Goal}
\newcommand{\sofid}{SO-Baseline}

\newcommand{\summnu}{\sum m_{\nu}}
\newcommand{\wde}{w_{0}}
\newcommand{\taure}{\tau_{\rm re}}
\newcommand{\omchsq}{\Omega_{c}h^{2}}
\newcommand{\ombhsq}{\Omega_{b}h^{2}}

\newcommand{\fskywideclean}{50\%}

\newcommand{\fskywidefull}{67\%}
\newcommand{\fskydeepfull}{3\%}
\newcommand{\fskyso}{40\%}
\newcommand{\fskysptpol}{1.2\%}
\newcommand{\fskysptsz}{6\%}
\newcommand{\fskysptthreeg}{3.6\%}

\newcommand{\howmanysimulations}{500}

\newcommand{\fsky}{f_{\rm sky}}

\newcommand{\boxsize}{$120^{\prime} \times 120^{\prime}$}
\newcommand{\pixres}{0.^{\prime}5}
\newcommand{\ukam}{\ensuremath{\mu}{\rm K{\text -}arcmin}}
\newcommand{\uk}{\ensuremath{\mu}{\rm K}}

\newcommand{\msol}{\ensuremath{\mbox{M}_{\odot}}}
\newcommand{\munits}{\times 10^{14}~\msol}

\newcommand{\lcdm}{$\Lambda$CDM}

\newcommand{\comment}[1]{}

\newcommand{\Mminredefine}{10^{13}}

\newcommand{\zmin}{0.1}
\newcommand{\zmax}{4.0}

\newcommand{\snrlimit}{5}
\newcommand{\alphay}{\alpha_{_{Y}}}
\newcommand{\betay}{\beta_{_{Y}}}
\newcommand{\gammay}{\gamma_{_{Y}}}
\newcommand{\alphasigmalogy}{\alpha_{\sigma}}
\newcommand{\gammasigmalogy}{\gamma_{\sigma}}
\newcommand{\YszM}{Y_{_{\rm SZ}}-M}
\newcommand{\hsebias}{b_{_{\rm HSE}}}
\newcommand{\hsebiasval}{0.2}

\newcommand{\thetamax}{2^{\prime}}
\newcommand{\rvir}{R_{\rm 500c}}

\newcommand{\mvir}{M_{\rm 500c}}
\newcommand{\Ysz}{Y_{\rm SZ}}
\newcommand{\Yvir}{Y_{{\rm SZ}_{500c}}}

\newcommand{\binnedcluscountsquantities}{z, M_{L}, q}

\newcommand{\binnedcluscounts}{N(\binnedcluscountsquantities)}

\newcommand{\clyy}{C_{\ell}^{yy}}
\newcommand{\clinv}{{\bf C}_{\rm \ell}^{-1}}
\newcommand{\bahamas}{\textsc{bahamas}}

\newcommand{\Ysymbol}{y}
\newcommand{\recoveredYsymbol}{\hat{\Ysymbol}}
\newcommand{\Ysymbolbold}{\boldsymbol{\Ysymbol}}
\newcommand{\recoveredYsymbolbold}{\boldsymbol{\recoveredYsymbol}}
\newcommand{\gautszcovsuppressionfactor}{\alpha}

\defcitealias{raghunathan21}{R21}

\newcommand{\ellnorm}{\ell_{\rm norm} = 3000}

\newcommand{\szpowerreduccmbhd}{\times 5}
\newcommand{\szpowerreducsfourdeep}{\times 3.1}
\newcommand{\szpowerreducsfoursptthreeg}{\times 2.4}
\newcommand{\szpowerreducsogoalsptpol}{\times1.5}

\newcommand{\szpowerreducsogoal}{\times 1.51}
\newcommand{\szpowerreducsptpol}{\times 1.49}
\newcommand{\szpowerreducsofid}{\times 1.35}
\newcommand{\szpowerreducsptsz}{\times 1.08}

\newcommand{\sfourdeepsensitivityimprovement}{3\%}
\newcommand{\cmbhdsensitivityimprovement}{12\%}

\newcommand{\sfourdeeptsznoisefloor}{10\%}
\newcommand{\cmbhdtsznoisefloor}{30\%}

\newcommand{\shortauthourlist}{Srinivasan Raghunathan}

\newcommand{\authourlist}
{
\author[0000-0003-1405-378X]{Srinivasan Raghunathan}
\email{srinirag@illinois.edu}
\affiliation{Center for AstroPhysical Surveys, National Center for Supercomputing Applications, Urbana, IL 61801, USA}
}

\newcommand{\abstracttext}{
We explore the significance of noise from thermal Sunyaev-Zel{'}dovich (tSZ) signals for cluster detection using cosmic microwave background (CMB) surveys. 
The noise arises both from neighboring objects and also from haloes below the detection limit. 
A wide range of surveys are considered: \sptsz, \sptpol, and \sptthreeg{} from the South Pole Telescope; \sofid{} and \sogoal{} configurations for Simons Observatory; CMB-S4's wide area (\sfour) and deep (\sfourdeep) surveys; and the futuristic \cmbhd{} experiment. 
We find that the noise from tSZ signals has a significant impact on \cmbhd{} and to some extent on \sfourdeep. 
For other experiments, the effect is negligible as the noise in the tSZ map is dominated by residual foregrounds or experimental noise. 
In the limit when the noise from tSZ signals is important, we find that removing the detected clusters and rerunning the cluster finder allows us to find a new set of less massive and distant clusters. 
Since the detected clusters are the dominant source of the tSZ power, removing them reduces the power at $\ell = 3000$ by: 
$\szpowerreduccmbhd$ for \cmbhd; $\szpowerreducsfourdeep$ of \sfourdeep; $\szpowerreducsfoursptthreeg$ for \sfour{} and \sptthreeg; $\szpowerreducsogoalsptpol$ for \sogoal{} and \sptpol; $\szpowerreducsofid$ for \sofid; and $\szpowerreducsptsz$ for \sptsz. 
We forecast the expected number of clusters and also derive parameter constraints by combining cluster counts with primary CMB and tSZ power spectra finding that the future surveys can reduce the error on the dark energy equation of state parameter to sub-percent levels and can also enable $\ge3\sigma$ detection of the sum of neutrino masses.
\ifdefined\ApJsubmit
\else
The simulation products and results can be downloaded from this \href{https://github.com/sriniraghunathan/tSZ_cluster_forecasts}{link$^{\text{\faGithub}}$}.
\fi
}

\newcommand{\tittext}{Assessing the Importance of Noise from Thermal Sunyaev-Zel{'}dovich Signals for CMB Cluster Surveys and Cluster Cosmology}

\begin{document}

\title{\tittext}
\shorttitle{Impact of {\rm tSZ}-noise on CMB Cluster Surveys}

\shortauthors{\shortauthourlist}
\authourlist

\begin{abstract}
\abstracttext{}
\end{abstract}


\section{Introduction}
\label{sec_introduction}

The thermal Sunyaev-Zel{'}dovich (tSZ) effect \citep{sunyaev70} is a result of the inverse Compton scattering of cosmic microwave background (CMB) photons off free electrons in the hot intracluster medium (ICM). 
The signal has emerged as a strong probe of both the structure formation as well as the complex ICM gastrophysics. 
For example, see works by \citet{komatsu02, shaw00, battaglia12, reichardt12, hill13b, horowitz17, bolliet18, douspis21} for astrophysical and cosmological constraints using the angular power spectrum of the tSZ and \citet{holder07, bhattacharya12, hill13, crawford14, hurier17} for statistics other than tSZ power spectrum. 
Besides these, the tSZ effect is also one of the most efficient methods used for detecting galaxy clusters \citep[see catalogs from][]{bleem15, planckSZcat16, hilton20, huang20, bleem20}.
The abundance of clusters as a function of mass and redshift can also place tight constraints on the parameters that govern structure formation and geometry of the Universe \citep{holder01a, lima04, allen11, sartoris12, mak13} like, for example, dark energy equation of state $\wde$, normalization of the matter power spectrum $\sigma_{8}$, and the sum of neutrino masses $\summnu$ as demonstrated using data \citep[recently by][]{zubeldia19, bocquet19, planck20_cosmo, to21, costanzi21, mantz21, salvati21}.
Thanks to its redshift independent nature \citep{sunyaev70}, the tSZ effect allows us to detect clusters at high redshifts where the signal-to-noise (\snr) of other cluster observables like richness estimates or X-ray flux drop rapidly. 
As a result, the tSZ-selected cluster samples from CMB surveys are highly complementary to large catalogs of $z \lesssim 1.5$ clusters expected from the future optical, infrared, and X-ray surveys \citep{lsst09, laureijs11, erosita12}. 

With the improvement in noise levels of the current \citep{benson14, henderson16, bender18} and future \citep{SO18, cmbs4collab19, sehgal19} CMB experiments, the sample size of the tSZ-selected clusters is expected to increase several fold compared to the existing catalogs \citep{louis17, madhavacheril17, gupta20, raghunathan21}.
However, as we are approaching these unprecedented map depths, the effect of astrophysical foreground signals also become important. 
Although the frequency dependent nature of the foreground signals can help us mitigate them using multi-frequency information,
one of the irreducible components is the noise from tSZ signals of other haloes, both along the line-of-sight (LOS) and also the confusion noise from the ones below the detection limit of the experiment. 
We collectively refer to both these sources as the tSZ-noise in the rest of the paper. 
\citet{holder07} explored the impact of the tSZ-noise and found it to be important for future surveys when the mass thresholds approach $M_{vir} \sim 1.2 \times 10^{14} h^{-1} \msol$. 
The authors, however, did not include the effects of other sources of variance like the astrophysical foregrounds, CMB, and the experimental noise in their study. 

In this work, we reassess the importance of the tSZ-noise for a wide range of CMB (past, current, and future) experiments namely: \sptsz{}, \sptpol{}, and \sptthreeg{} surveys from the South Pole Telescope (SPT) and the upcoming experiments namely Simons Observatory (\simonsobs), \cmbsfour{}, and \cmbhd{}.
For \simonsobs, we consider two configurations: \sofid{} and \sogoal{} \citep{SO18}.
For \cmbsfour{}, we use both the wide-area survey (\sfour) from Chile and the smaller but deeper delensing survey (\sfourdeep) from the South Pole \citep{cmbs4collab19}. 
Our simulations include the effect of experimental noise and the astrophysical foregrounds. 
We find that the tSZ-noise has no impact on cluster surveys from \sptsz{}, \sptpol{}, \sofid{} and \sogoal; and a negligible impact on \sptthreeg{} and \sfour. 
The effect of tSZ confusion noise is mildly important for \sfourdeep{} but significantly affects \cmbhd. 
For all surveys, we also show the impact of masking the detected clusters in the tSZ power spectrum.  
When the tSZ-noise affects cluster finding, we find that removing the detected clusters and rerunning the cluster detection algorithm allows us to detect less massive and distant clusters lying near the same LOS. 
Finally, we compute the expected number of clusters and the cosmological constraints by combining cluster abundance measurements with primary CMB and tSZ power spectra. 

The paper is structured as follows. 
In \S\ref{sec_sims}, we describe our simulations: CMB and astrophysical foregrounds in \S\ref{sec_cmb_astrofg}, noise in \S\ref{sec_noise_model}, tSZ signals in \S\ref{sec_diffusetsz} and \S\ref{sec_clustertsz}. 
The cluster detection method is given in \S\ref{sec_cluster_detection}. 
We present the formalism to get cluster counts and Fisher forecasts is in \S\ref{sec_fisher_formalism}, results in \S\ref{sec_results},  and conclude in \S\ref{sec_conclusion}. 
Finally, we provide details about the cluster $\snr$ calculation in  Appendix~\ref{appendix_snr_calculation}. 
 
The cluster masses in this work are reported with respect to the radius $\rvir$ which encompasses the region within which the average mass density is 500 times the critical density of the Universe at the cluster redshift $z$. 
We adopt the flat-sky approximation and replace spherical harmonic transforms by Fourier transforms since we are dealing with small regions of the sky.
The relation between the Fourier wavenumber $k$ and the multipole $\ell$ is given by $k = \sqrt{k_x^{2} + k_y^{2}} = \ell/2\pi$. 
The underlying cosmology used was set to \planck{} 2015 measurements (TT + lowP in Table 4 of \citealt{planck15-13}).

\section{Simulations}
\label{sec_sims}
In this section, we briefly describe the simulations and cluster detection method used in this work and point the reader to \citet[][referred to as  \citetalias{raghunathan21} hereafter]{raghunathan21} for more details.
The simulation setup is similar to \citetalias{raghunathan21} other than the tSZ simulations and noise models for experiments not considered in that work.
 
\subsection{CMB and astrophysical foregrounds}
\label{sec_cmb_astrofg}

We generate \boxsize{} wide realizations of the millimetre wave temperature sky with a pixel resolution of $\pixres$. 
The simulated components include: CMB; instrumental and atmospheric noise; astrophysical foregrounds namely emission from radio point sources, dusty star-forming galaxies (DSFG) which we refer to as the cosmic infrared background (CIB), diffuse kinematic SZ (kSZ) and tSZ signals, and the tSZ signal from the cluster under study. 
For simplicity, we ignore signals from galactic foregrounds but note that they can affect large-scale Compton-$y$ measurements and hence have non-negligible impact only on massive low redshift cluster detections \citepalias{raghunathan21}. 
The simulated maps are convolved with a band-dependent Gaussian instrumental beam. The instrumental specifications and the number of frequency bands for the experiments are given in Table~\ref{tab_exp_specs}. 
We note that the pixel resolution adopted here is sub-optimal for the \cmbhd{} experiment and hence our results must be treated as conservative estimates. 
We perform Gaussian simulations of an underlying power spectrum for CIB, CMB, kSZ, noise and radio galaxies. 
For tSZ, we use a different approach as described in \S\ref{sec_diffusetsz}.

To simulate the CMB, we use \texttt{CAMB} \citep{lewis00} software to obtain the LSS lensed power spectra $C_{\ell}^{TT}$ for the fiducial \planck{} 2015 cosmology. 
Modeling of the atmospheric and instrumental noise spectra is described in \S\ref{sec_noise_model}.
For astrophysical foregrounds namely CIB, kSZ, and radio galaxies, we use power spectra measured at 150 GHz by the SPT experiment \citep{george15} and simply scale them to other bands. 
Note that the power from CIB, kSZ, and radio galaxies are dominated by the diffuse signals below the detection limit and as a result performing Gaussian realizations is a reasonable thing to do.
To scale the spectrum of the radio galaxies for sources with flux $S_{150}<$ 6.4 mJy which is the SPT masking threshold \citep{george15, reichardt21}, we use a power law with spectral index $\alpha_{\rm radio} = -0.6$ \citep{everett20}.
For CIB, we use the spectral energy distribution (SED) scaling parameterized using a modified blackbody as $\eta_{\nu} = \nu^{\beta_{\rm CIB}}\ B_{\nu}(T_{\rm CIB})$ where $T_{d}$ = 20 K is the temperature and $B_{\nu}(T)$ is the \planck{} function. 
We use two different emissivity indices $\beta_{\rm CIB-PO} = 1.505$ and $\beta_{\rm CIB-Cl} = 2.51$ for the Poisson and the clustered components of the CIB signal respectively \citep{george15}. 
The masking threshold used for DSFG is the same as radio galaxies ($S_{150}=$ 6.4 mJy) for all experiments except \cmbhd.
For \cmbhd, we follow \citet{sehgal19} and reduce the level of CIB by $\times17$ at 150 GHz. 
This is to account for the template based subtraction of sources with flux above $S_{150} \ge$ 0.04 mJy using higher $\snr$ detections of the same sources at 270/350 GHz bands \citep{sehgal19}. 
Since the kSZ has the same blackbody spectrum as the CMB, we do not apply any frequency scaling for the diffuse kSZ signals. 

\begin{deluxetable*}{| l | c | ccccccc | ccccccc |}[h]
\def\arraystretch{1.2}
\tablecaption{Instrumental specifications for different experiments considered in this study.}
\label{tab_exp_specs}
\tablehead{
\multirow{2}{*}{Experiment} & \multirow{2}{*}{$f_{\rm sky}$} & \multicolumn{7}{c|}{Beam $\theta_{\rm FWHM}$ [arcminutes]} & \multicolumn{7}{c|}{$\Delta_{T}$ [$\ukam$]}\\
\cline{3-16}
& & 30 & 40 & 90 & 150 & 220 & 270 & 350 & 30 & 40 & 90 & 150 & 220 & 270 & 350
}
\startdata
\sptsz{} & \fskysptsz & - & -  & 1.7 & 1.2 & 1.0 & -  & - &  - & -  & 41.0 & 18.0 & 80.0 & - & - \\\hline
\sptpol{} & \fskysptpol & -  & -  & 1.7  & 1.2 & - & - & - & - & - & 12.0 & 5.5 & - & - & - \\\hline
\sptthreeg{} & \fskysptthreeg & - & -  & 1.7 & 1.2 & 1.0 & -  & - & -  & - & 3.0 & 2.2 & 8.8 & - & - \\\hline\hline
\sofid{} & \fskyso & 7.4 & 5.1 & 2.2 & 1.4 & 1.0 & 0.9 & - & 71.0 & 36.0 & 8.0 & 10.0 & 22.0 & 54.0 & - \\\hline
\sogoal{} & \fskyso & 7.4 & 5.1 & 2.2 & 1.4 & 1.0 & 0.9 & - & 52.0 & 27.0 & 5.8 & 6.3 & 15.0 & 37.0 & - \\\hline\hline
\sfour{} & \fskywidefull & 7.3 & 5.5 & 2.3 & 1.5 & 1.0 & 0.8 & - & 21.8 & 12.4 & 2.0 & 2.0 & 6.9 & 16.7 & - \\\hline
\sfourdeep{} & \fskydeepfull & 8.4 & 5.8 & 2.5 & 1.6 & 1.1 & 1.0 & - & 4.6 & 2.94 & 0.45 & 0.41 & 1.29 & 3.07 & - \\\hline\hline
\cmbhd{} & \fskywideclean & 1.4 & 1.05 & 0.45 & 0.25 & 0.20 & 0.15 & 0.12 & 6.5 & 3.4 & 0.73 & 0.79 & 2.0 & 2.7 & 100 \\\hline\hline
\enddata
\end{deluxetable*}

\subsection{Noise models}
\label{sec_noise_model}

\begin{deluxetable}{| c | c | c | c | c |}[t]
\tabletypesize{\footnotesize}
\tablecaption{Atmospheric $1/f$ noise specifications ($\ell_{\rm knee}$, $\alpha_{\rm knee}$) for  \spt{} (\sptthreeg, \sptpol, and \sptthreeg),  \cmbsfour, and \cmbhd{} experiments. For \sptpol, we do not include the 220 GHz band in line with Table~\ref{tab_exp_specs}.}
\label{tab_exp_atm_noise_cmbs4_spt}
\tablehead{
Band [GHz] & \spt\tablenotemark{a} & \sfourdeep\tablenotemark{b} & \sfour\tablenotemark{c} & \cmbhd\tablenotemark{c}}
\startdata
30 & \multirow{2}{*}{-} & 1200, 4.2 & \multicolumn{2}{c|}{471, 3.5} \\
\cline{1-1}\cline{3-5}
40 & & 1200, 4.2 & \multicolumn{2}{c|}{478, 3.5} \\\hline
90 & 1200, 3.0 & 1200, 4.2 & \multicolumn{2}{c|}{2154, 3.5} \\\hline
150 & 2200, 4.0 & 1900, 4.1 & \multicolumn{2}{c|}{4364, 3.5} \\\hline
220 & 2300, 4.0 & 2100, 3.9 & \multicolumn{2}{c|}{7334, 3.5} \\\hline
270 &  \multirow{2}{*}{-} &  2100, 3.9 & \multicolumn{2}{c|}{7308, 3.5} \\
\cline{1-1}\cline{3-5}
350 & & - & -  & 7500, 3.5\\\hline\hline
\enddata
\tablenotetext{a}{Check Appendix of \citet{aylor19}.}
\tablenotetext{b}{Check V3R025 configuration in \url{https://cmb-s4.uchicago.edu/wiki/index.php/Delensing_sensitivity_-_updated_sensitivities,_beams,_TT_noise}.}
\tablenotetext{c}{\url{https://cmb-s4.uchicago.edu/wiki/index.php/Expected_Survey_Performance_for_Science_Forecasting}.}
\end{deluxetable}

Experimental noise can be decomposed into two: atmospheric and instrumental noise components. 
The combined noise power spectrum is modeled as 
\begin{equation}
\label{eq_noise_model}
N_{\ell} = N_{\rm \ell, white} + N_{\rm \ell, atm} \left( \frac{\ell}{\ell_{\rm knee}}\right)^{-\alpha_{\rm knee}}.
\end{equation}
The first term $N_{\rm \ell, white}$ represents the instrumental noise as given in Table~\ref{tab_exp_specs} for different experiments \citep{bleem15, henning18, bender18, SO18, cmbs4collab19, sehgal19}. 
The rest of the terms, $N_{\rm \ell, atm}, \ell_{\rm knee}$, and $\alpha_{\rm knee}$ are used to model the atmospheric noise.
For \cmbhd, \cmbsfour{} and \spt{} (\sptthreeg, \sptpol, and \sptsz) experiments we set $N_{\rm \ell, atm} = N_{\rm \ell, white}$ and the values used for $\ell_{\rm knee}$ and $\alpha_{\rm knee}$ are provided in Table~\ref{tab_exp_atm_noise_cmbs4_spt}. 
For \simonsobs, we adopt the same prescription given in \citet{SO18} by using $N_{\rm \ell, red}$ and setting $\ell_{\rm knee} = 1000$ and  $\alpha_{\rm knee}=3.5$ \citep{louis16}.
\newcommand{\soNredunits}{\uk^{2}s\ {\rm Sr}^{-1}}
The values for $N_{\rm \ell, red}$ in $\soNredunits$ for \simonsobs{} large aperture telescopes (LAT) are picked from Table 2 of \citet{SO18}. 
We convert $N_{\rm \ell, red}$ in $\soNredunits$ to $N_{\rm \ell, atm}$ in $\uk^{2}$ units using the procedure detailed in \S2.2 of \citet{SO18} as 
\begin{eqnarray}
N_{\rm \ell, atm} = N_{\rm \ell, red} \dfrac{A_{\rm sky}}{t_{\rm obs}}. 
\end{eqnarray}
Here the sky area is $A_{\rm sky} = 4\pi \fsky\ [\rm Sr]$ and $\fsky = 0.4$ as given in Table~\ref{tab_exp_specs}.
The total observation time is \mbox{$t_{\rm obs} = N_{\rm total}\ \eta_{\rm obs-eff}\ \eta_{\rm map-cuts}$} where $N_{\rm total}$ = 5 years is the survey period expressed in seconds, $\eta_{\rm obs-eff} = 0.2$ is the observing efficiency, and $\eta_{\rm map-cuts} = 0.85$ represents the data cuts to account to $\sim 15\%$ noisy map edges \citep{SO18}.

\subsection{Thermal SZ simulations}
\label{sec_diffusetsz}

\begin{figure}
\centering
\ifdefined\ApJsubmit
\includegraphics[width=0.48\textwidth, keepaspectratio]{poisson_sims_all_haloes.pdf}
\else
\includegraphics[width=0.48\textwidth, keepaspectratio]{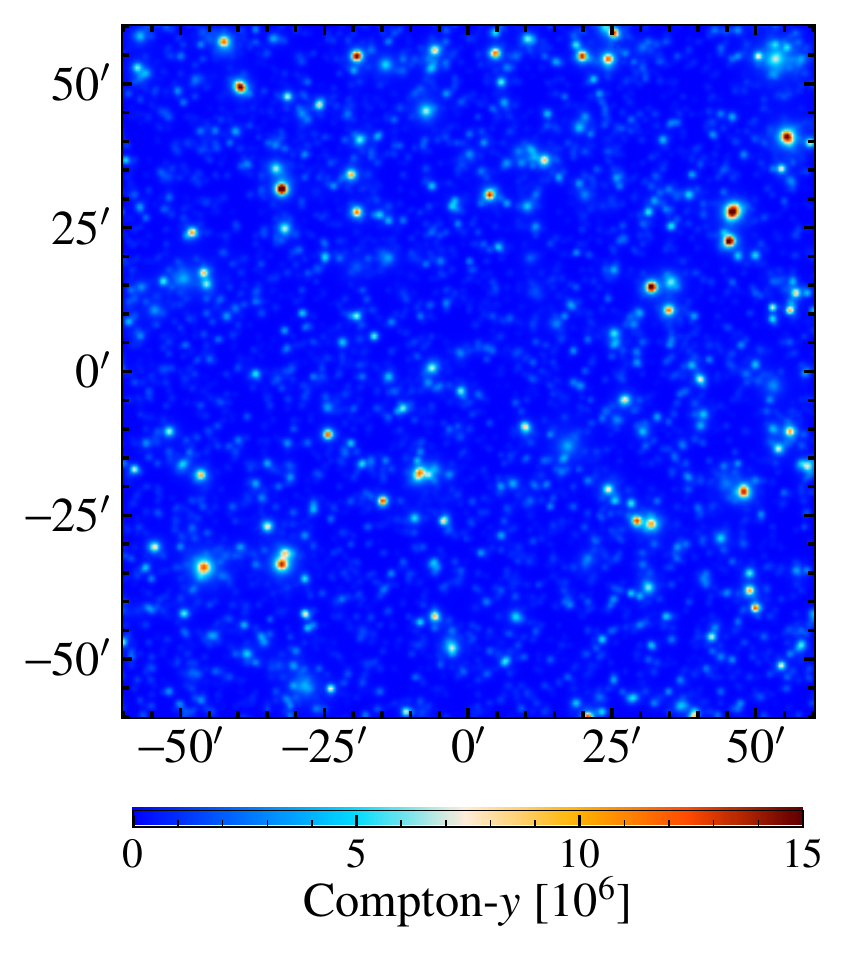}
\fi
\caption{An example of the \boxsize{} Poisson tSZ simulation produced in this work. 
The simulation contains signal from all clusters with $\mvir \ge \Mminredefine\ \msol$ in the redshift range $z \in [\zmin, \zmax]$. 
The tSZ signal of a given halo is modeled using the generalized NFW profile as described in \S\ref{sec_clustertsz} and the signal arising due to clustering of clusters has been ignored.
}
\label{fig_poisson_tsz_simulation}
\end{figure}

The main goal of this work is to understand the importance of the noise from the tSZ signals of haloes other than the cluster under consideration. 
Like mentioned earlier, the tSZ-noise can arise both from adjacent haloes near the LOS and also from the haloes below the detection limit of the experiment. 
While it is reasonable to perform Gaussian realizations of the diffuse tSZ signals below the detection threshold, assuming the same for all haloes can degrade the cluster detection sensitivity. 
This is because the number counts of clusters follow a random Poisson distribution on the sky, which is highly non-Gaussian, and the total map variance is dominated by a few pixels with clusters. 
Here and throughout, when we refer to non-Gaussian nature of the tSZ signals, we simply refer to the non-Gaussian tails in the Poisson distribution and not to any other higher point correlations, which are assumed to be absent in this study. 
Subsequently, we generate Poisson realizations of the \boxsize{} tSZ sky by considering all haloes with $\mvir \ge \Mminredefine\ \msol$ in the redshift range $z \in [\zmin, \zmax]$. 

\begin{figure}
\ifdefined\ApJsubmit
\includegraphics[width=0.48\textwidth, keepaspectratio]{compton_y_powerspec_fiducial.pdf}
\else
\includegraphics[width=0.48\textwidth, keepaspectratio]{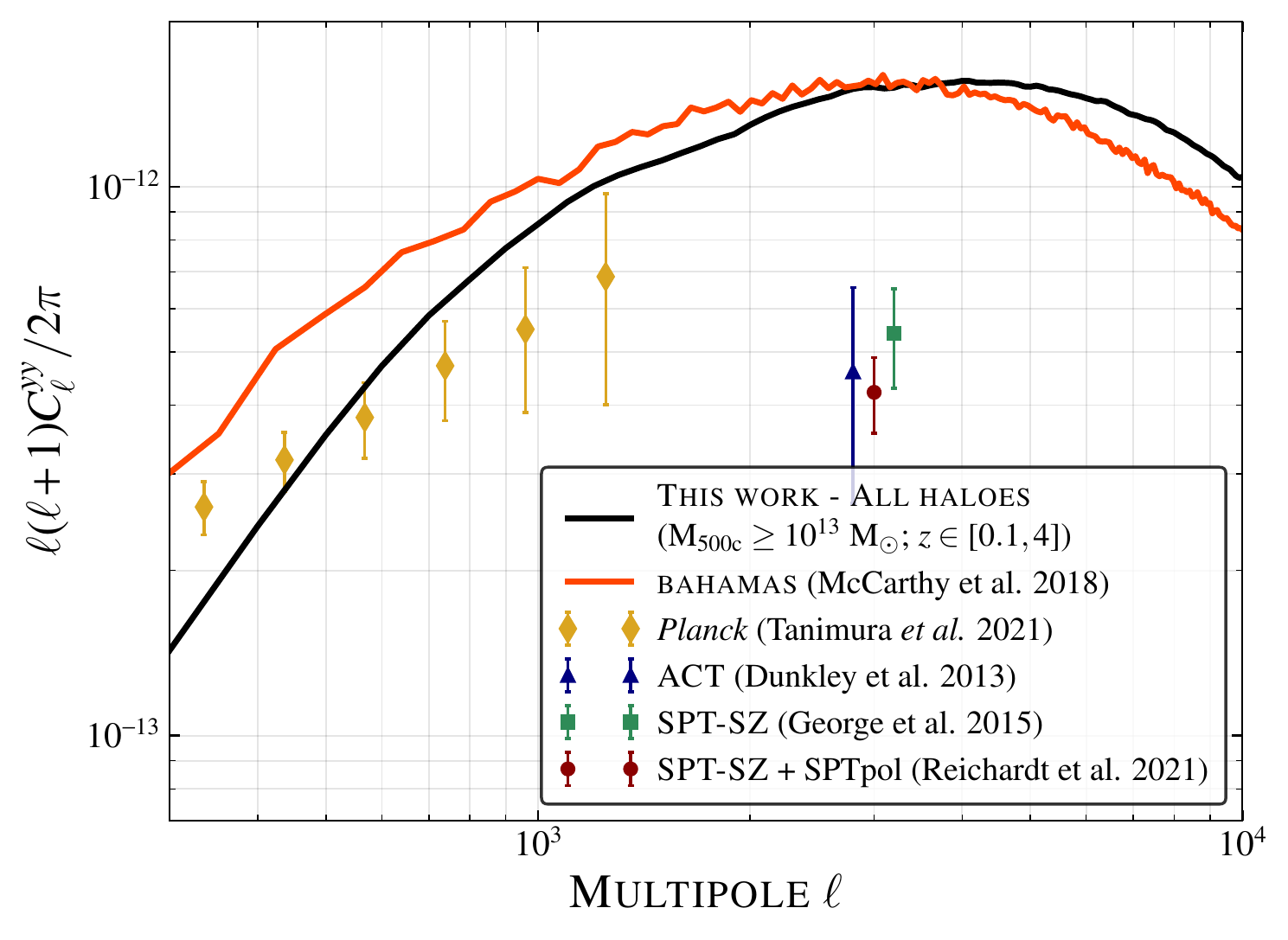}
\fi
\caption{The black solid curve is the average Compton-$y$ power spectrum, expressed as $D_{\ell}^{yy}$, from \howmanysimulations{} Poisson simulations generated in this work. 
The orange curve is the power spectrum computed using the publicly available \bahamas{} simulations for the fiducial model \citep{mccarthy18}. 
For reference, we also show the measurements from \planck{} as yellow diamonds \citep{plancksz15}, ACT as blue triangle \citep{dunkley13}, \sptsz{} and \sptpol{} as green square \citep{george15} and red circle \citep{reichardt21}. 
}
\label{fig_poisson_tsz_simulation_power_spectra}
\end{figure}

To this end, we adopt the  \citet{tinker08} halo mass function (HMF) to first generate a map of the number counts and paste cluster tSZ signal of the corresponding halo $y(\mvir, z)$ as described below in \S\ref{sec_clustertsz}.
The simulations do not include contributions from clustering of clusters, the {\it 2-halo term}, since it has been found to be negligible in the angular scales \citep[see Fig. 1 of][]{komatsu99} and mass range \citep[see Fig. 1 and Fig. 7 of][]{hill18} considered in this work. 
However, we also note that this is strictly true only for tSZ power spectrum measurements since the presence of clustering can increase the local variance and might have an impact on the cluster detection. We leave a careful investigation of this to a future study.
Fig.~\ref{fig_poisson_tsz_simulation} shows an example of the Poisson-tSZ simulation generated in this study. 
The average Compton-$y$ power spectrum from \howmanysimulations{} Poisson realizations is shown as the black solid curve in Fig.~\ref{fig_poisson_tsz_simulation_power_spectra}. 
While our simulations predict a higher power compared to the values reported by ACT, \planck{}, and SPT, the power spectrum is in good agreement with the publicly available \bahamas{}\footnote{We pick the fiducial \bahamas{} model for the \planck{} 2015 cosmology from \url{https://www.astro.ljmu.ac.uk/~igm/BAHAMAS/}} simulations \citep{mccarthy17, mccarthy18} shown as the orange solid curve. 
We show the measured values by ACT \citep{dunkley13} as blue triangle, \planck{} \citep{tanimura22} as yellow diamonds, \sptsz{} \citep{george15} as green square, and the combination of \sptsz{} and \sptpol{} \citep{reichardt21} as red circle.
The fiducial \bahamas{} model shown here includes astrophysical feedback which pushes the gas out from the haloes reducing the power on small-scales compared to the case without feedback effects \citep{shaw10, mccarthy14}. 
Since our simulations 
do not include such astrophysical effects, the black curve is higher than the orange \bahamas{} model on small-scales. 
Note that we do not perform any normalization using \bahamas{} and the match between black and orange curves around $\ell \sim 3000$ is just a coincidence. 
On large-scales, the power spectrum is slightly lower than \planck{} and \bahamas{} primarily because of the minimum cluster redshift $z = 0.1$ adopted in this work but also due to absence of the clustering term.
We ignore such large-scale effects since those scales are less important for clusters expected from future CMB surveys. 

Once the tSZ simulations are produced, we inject the tSZ signals using three different approaches in our simulated maps. 
In the first case, we simply add the generated Poisson realizations (see Fig.~\ref{fig_poisson_tsz_simulation} for an example) and this is referred as Poisson-tSZ case throughout this work. 
In the second (Gaussian-tSZ) case, we use the fiducial power spectrum shown as black solid curve in Fig.~\ref{fig_poisson_tsz_simulation_power_spectra} and produce Gaussian realizations of them.
The difference in the results between the above two cases shows the importance of the non-Gaussian nature of the tSZ. 
We also compare the results from Poisson-tSZ case to the results from simulations where the tSZ signals are completely ignored (no-tSZ). 
This quantifies the impact of the tSZ-noise on cluster detection.

\subsection{Cluster tSZ signal}
\label{sec_clustertsz}

The cluster tSZ signal is modeled using a generalized Navarro-Frenk-White (NFW, \citealt{navarro96, zhao96}) as proposed by \citet{nagai07} and calibrated using X-ray observations by \citet{arnaud10}. 
The dimensionless pressure profile $P_{e}(l, x)$ of the ICM is integrated along the LOS to obtain the Compton-$y$ signal $y(x)$ which is then converted into  CMB temperature units as 
\begin{eqnarray}
\label{eq_pressure_compton_y}
P_{e}(l, x) &  = &  \dfrac
{ P_{0} }
{ 
(c_{500}x)^{\gamma} \left[1 + (c_{500}x)^{\alpha}\right]^{\left(\frac{\beta-\gamma}{\alpha} \right)}
}\,\\\notag
y(x) & = &  \frac{\sigma_{T}}{m_{e} c^{2}} \int_{l}P_{e}(l, x)\ dl\,\\\notag
\delta_{T} & = &  y(x)\ g_{\rm SZ}(\nu) T_{\rm CMB}\ {\rm K}.
\end{eqnarray}
The frequency dependence of the tSZ signal \citep{itoh98, chluba12} is 
\begin{eqnarray}
\label{eq_tsz_freq_dep}
g_{\rm SZ}(\nu) = x\ {\rm coth}(x/2) - 4\ {\rm with}\ x = \frac{h \nu}{k_{B}T_{\rm CMB}}.
\end{eqnarray}
In the above equations, $c_{500}x = r/r_{s}$ where $c_{500}$ is the concentration parameter, $r_{s}$ is the scale radius, and $x = r / R_{500}$ is the distance to cluster center expressed in terms of $R_{500}$;
$c$ in the velocity of light, $m_{e}$ is the mass of the electron, $\sigma_{T}$ in the Thomson cross section, $T_{\rm CMB} = 2.73\ {\rm K}$ is the mean temperature of the CMB; $h$ and $k_{B}$ are the Planck and Boltzmann constants respectively.
Using the generalization of the \citet{plancksz15} $\YszM$ relation by \citet{louis17, madhavacheril17} to include mass and redshift evolution, we obtain the integrated cluster Compton $\Yvir$ by integrating $y(x)$ over the angular extent of the cluster $R_{500}$ as 

\begin{eqnarray}
\label{eq_Ysz_mass}
\Yvir = Y_{\ast}\ \left[\dfrac{H_{0}}{70}\right]^{-2+\alphay}\ \left[ \dfrac{(1-\hsebias) \mvir}{M_{\ast}} \right]^{\alphay} \\\notag
{\rm e}^{\betay {\rm log}^{2} \left( \frac{\mvir}{M_{\ast}} \right)}\ \left[\dfrac{D_{A}(z)}{100 {\rm Mpc}}\right]^{2} E^{2/3}(z)\  (1+z)^{\gammay},
\end{eqnarray}
along with a log-normal scatter 
\begin{eqnarray}
\label{eq_Ysz_mass_scatter}
\sigma_{{\rm log}Y}(\mvir, z) = \sigma_{{\rm log}Y, 0} \left[ \dfrac{ \mvir}{M_{\ast}} \right]^{\alphasigmalogy} (1 + z)^{\gammasigmalogy}.
\end{eqnarray}
We adopt the same pivotal mass $M_{\ast} = 6 \munits$ as \citet{plancksz15} and subsequently set the normalization factor ${\rm log}Y_{\ast} = -0.19$ and the mass evolution parameter $\alphay = 1.79$. 
The hydrostatic equilibrium (HSE) mass bias parameter is assumed to be constant across all redshifts and set to $\hsebias = \hsebiasval$ \citep{zubeldia19, ryu20}. 
The angular diameter distance to the cluster at redshift $z$ and the Hubble function are represented using $D_{A}(z)$ and $E(z) = H(z)/H_{0}$ respectively. 
The fiducial value for the scatter is set to $\sigma_{{\rm log}Y, 0} = 0.127$ \citep{louis17, madhavacheril17}. 
The second order mass evolution $\betay$ and redshift evolution $\gammay$ parameters of the $\YszM$ relation in Eq.(\ref{eq_Ysz_mass}) and the mass $\alphasigmalogy$ and redshift $\gammasigmalogy$ evolution parameters of the scatter Eq.(\ref{eq_Ysz_mass_scatter}) are all fixed to be zero.

\begin{figure*}
\centering
\ifdefined\ApJsubmit
\includegraphics[width=0.82\textwidth, keepaspectratio]{ilc_weights.pdf}
\else
\includegraphics[width=0.82\textwidth, keepaspectratio]{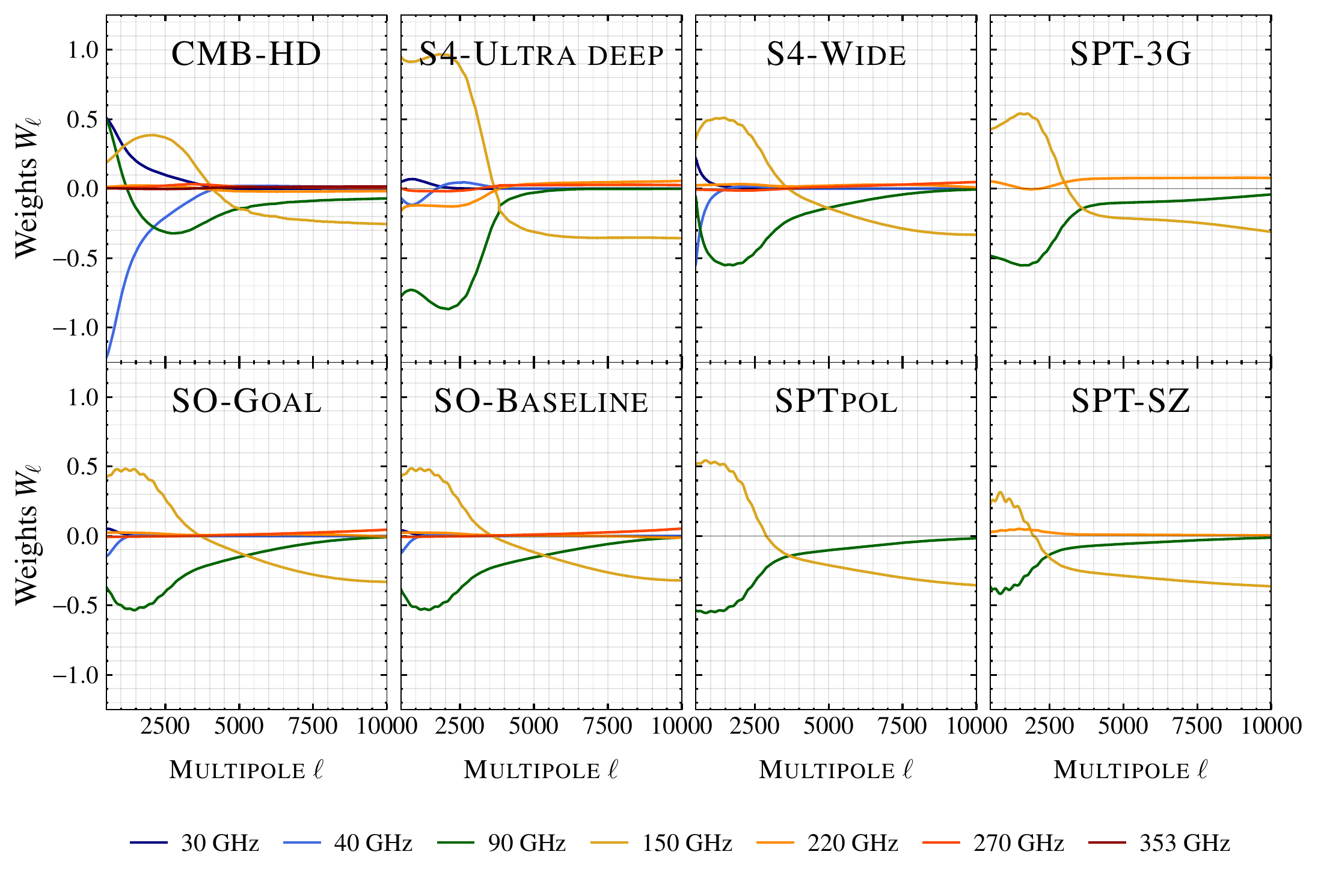}
\fi
\caption{Multipole dependent weights for different frequency bands used to construct the minimum variance Compton-$y$ map from different experiments considered in this work. 
The weights are dominated by 90 and 150 GHz channels. 
At large-scales, 30 and 40 GHz can add some information due to the reduced level of atmospheric noise. 
Bands higher than 220 GHz are primarily used to reduce the variance from CIB. 
}
\label{fig_ilc_weights}
\end{figure*}

\section{Cluster detection}
\label{sec_cluster_detection}

Similar to \citetalias{raghunathan21}, we perform an optimal internal linear combination (ILC) of maps $M$ from different frequency channels $N_{\rm ch}$ to construct a minimum variance (MV) Compton-$y$ map that is fed into a maximum likelihood estimator (MLE) for $\snr$ calculation.  
In Fourier space, the ILC operation corresponds to $y_{\ell} = \sum_{i=1}^{\rm N_{ch}} w_{\ell}^{i} M_{\ell}^{i}$. 
We obtain the optimal multipole dependent weights $w_{\ell}$ for each frequency channel using the \texttt{SMICA} (Spectral Matching Independent Component Analysis) algorithm \citep{cardoso08, remazeilles11, planck14_smica} as 
\begin{eqnarray}
w_{\ell} = \dfrac{\clinv {\bf a}}{{\bf a}^{T} \clinv {\bf a}}, 
\label{eq_ilc_weights}
\end{eqnarray}
where ${\bf C}_{\ell}$ is the covariance between the simulated maps in multiple frequencies at a given multipole $\ell$ with dimension $\rm N_{ch} \times \rm N_{ch}$, \mbox{${\bf a} = [-5.33, -5.23, -4.36, -2.61, 0.09, 2.27, 5.95]$} is the frequency response vector of the Compton-$y$ signal for [30, 40, 90, 150, 220, 270, 350] GHz channels shown in Eq.(\ref{eq_tsz_freq_dep}). 

In Fig.~\ref{fig_ilc_weights} we show the weights $w_{\ell}$ for all channels used to construct the MV Compton-$y$ map from different experiments. 
Since we are mainly interested in the tSZ-noise, we do not use constrained ILC algorithms \citep{remazeilles11} as done in \citet{madhavacheril20, bleem21}. 
For details about the effect of CIB / radio galaxy contamination on the recovered tSZ signals, we refer the reader to look into \citetalias{raghunathan21}.
We can note from the figure that 90 and 150 GHz channels mostly dominate the weights as expected. 
The weights for higher frequency channels ($\nu \ge$ 220 GHz) are significantly smaller since they are mainly used to reduce the variance from the CIB signals. The lower frequency bands, 30/40 GHz, have a much larger beam than 90/150 GHz channels and only contribute on large-angular scales to reduce the variance from atmospheric noise. 

\begin{figure}
\centering
\ifdefined\ApJsubmit
\includegraphics[width=0.48\textwidth, keepaspectratio]{ilc_residuals.pdf}
\else
\includegraphics[width=0.48\textwidth, keepaspectratio]{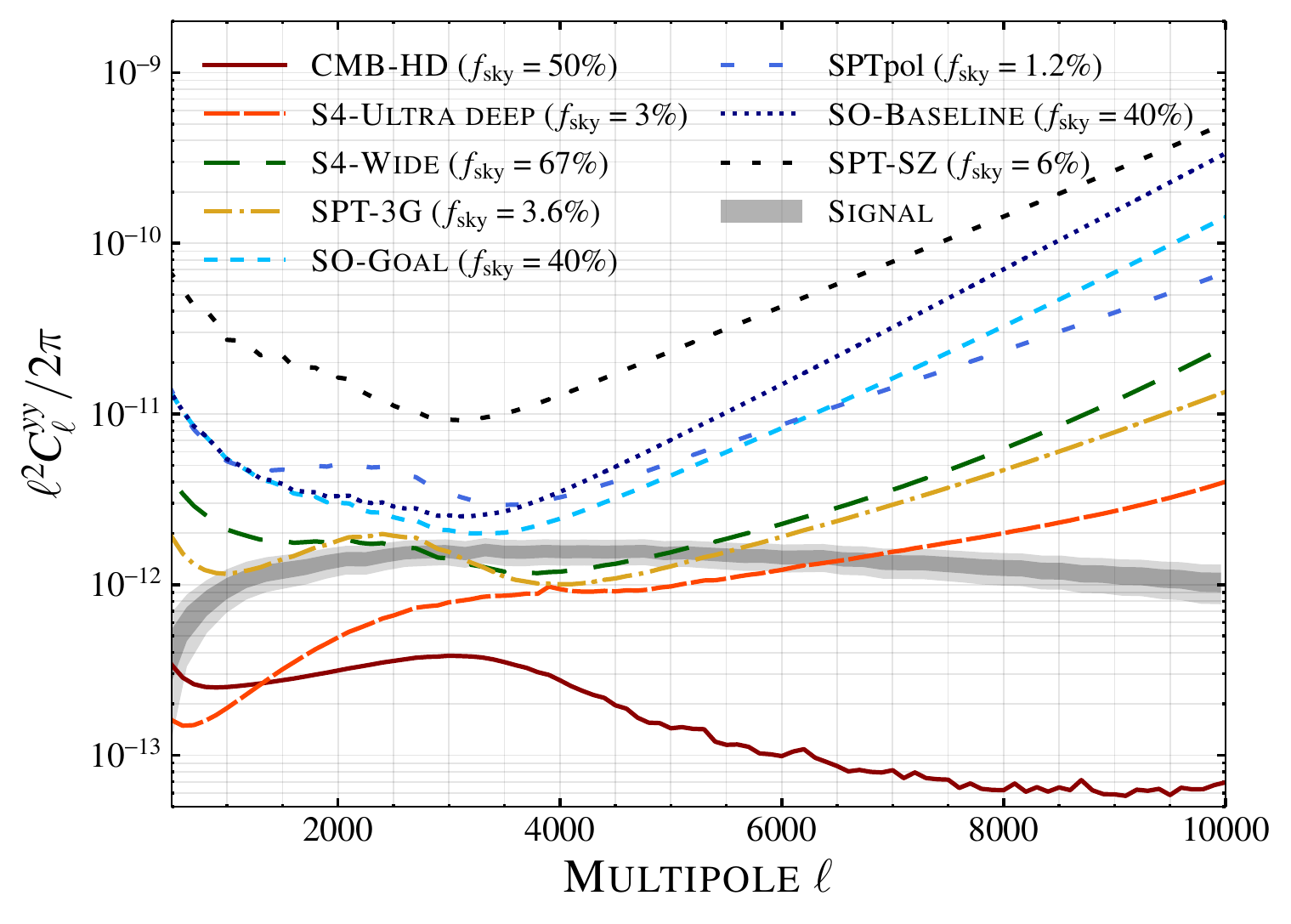}
\fi
\caption{
Noise residuals $N_{\ell}^{yy}$ in the Compton-$y$ maps from different experiments considered in this work. 
The grey band represents the fiducial signal level from simulations with $1\sigma, 2\sigma$ errors from the SPT-SZ survey \citep{george15}. 
\sptthreeg{} and \sfour{} will map the peak of the tSZ power spectrum with $\snr \ge 1$ and the $\snr$ for \sfourdeep{} and \cmbhd{} are much higher. 
The residual noise in the Compton-$y$ map for \cmbhd{} is lower than the fiducial signal level at $\ell \ge 500$ because of the reduced level of CIB signals. 
The small-scale noise in other experiments contains significant contributions from both CIB and instrumental noise. 
The sky area scanned by each experiment is also provided in the legend.
}
\label{fig_ilc_residuals}
\end{figure}

The sum of the weighted noise and foregrounds signals, $N_{\ell}^{yy} = \dfrac{1}{{\bf a}^{T} \clinv {\bf a}}$, referred generally as residual noise in the Compton-$y$ maps are presented in Fig.~\ref{fig_ilc_residuals}. 
The grey band in the figure represents the $1\sigma$ and $2\sigma$ errors from \citep{george15} and the fiducial value in the figure is set to mean value from \howmanysimulations{} Poisson simulations performed in this work (same as black solid curve in Fig.~\ref{fig_poisson_tsz_simulation_power_spectra}). 
On small scales, the residuals are much lower for \cmbhd{} compared to other experiments because of the reduced level of CIB signals (see \S\ref{sec_cmb_astrofg}) expected in the  \cmbhd{} maps \citep{sehgal19}. 
The other experiments are limited by foregrounds signals, primarily CIB, or instrumental noise. 
On large-scales, in the absence of galactic foregrounds as is the case here, atmospheric noise and CMB dominate the residuals. 
They can be lowered further by including information from the \planck{} satellite. 
From the figure, we can see that \sptthreeg{} and \sfour{} will be sample variance limited near the peak $3000 \le \ell \le 5000$ of the Compton-$y$ spectrum  and map those scales with $\snr \ge 1$. 
The residual noise in the Compton-$y$ maps for \sfourdeep{} and \cmbhd{} are much lower than the other experiments and the fiducial signal level. 
As a result, we can suppose the tSZ-noise to have a significant impact on the cluster survey from \cmbhd{} and to some extent from \sfourdeep{} compared to other experiments as we show in \S\ref{sec_baseline_results}. 

We feed the MV Compton-$y$ map into the MLE to estimate the $\snr$ of clusters as a function of $\mvir$ and redshift $z$. 
We calculate 
\begin{equation}
    \label{eq_likelihood}
    -2 \ln{} \mathcal{L} = \sum_{ij} \left(\recoveredYsymbolbold_i - \Ysymbolbold^\mathrm{th}_i \right) \mathbf{\hat{C}}_{ij}^{-1} \left(\recoveredYsymbolbold_j - \Ysymbolbold^\mathrm{th}_j \right)\,,
\end{equation}
where $\Ysymbolbold_{i} \equiv \Ysymbolbold_{i}(\theta)$ is the  azimuthally-averaged profile of the Compton-$y$ signal in bins $i$ of \mbox{$\Delta \theta = \pixres$}. 
The models $\Ysymbolbold^\mathrm{th}$ used for fitting the measured Compton-$y$ signal $\recoveredYsymbolbold$, are calculated using Eq.(\ref{eq_Ysz_mass}). 
We set $\theta_\mathrm{max}=\thetamax$ since 
$\theta_\mathrm{max} \ge  \rvir/D_{A}(z)$ encompasses the tSZ signal from majority of the clusters at all redshifts. 
Finally, the covariance matrix $\mathbf{\hat{C}}$ is computed using \mbox{$N = 2500$} simulations as
\begin{eqnarray}
    \label{eq_cov_matrix}
    \mathbf{\hat{C}} = \frac{1}{N-1}\sum\limits_{n = 1}^{N} \left(\recoveredYsymbolbold_i - \left< \recoveredYsymbolbold \right> \right) \left(\recoveredYsymbolbold_i - \left< \recoveredYsymbolbold\right> \right)^{T}\,.
\end{eqnarray}

We decompose $\mathbf{\hat{C}}$ into two components 
\begin{eqnarray}
    \label{eq_cov_matrix_components}
    \mathbf{\hat{C}} = \mathbf{\hat{C}}_{\rm other} + \mathbf{\hat{C}}_{\rm tSZ}.
\end{eqnarray}
The first term $\mathbf{\hat{C}}_{\rm other}$ is calculated using maps from different bands that contain CMB, noise, and all astrophysical foregrounds other than the tSZ after passing them through the ILC pipeline to get the MV Compton-$y$ map. 
This Compton-$y$ only contains all sources of noise but does not include the cluster tSZ or tSZ signals from other haloes.

As mentioned in \S\ref{sec_diffusetsz}, we inject the tSZ signals using three different approaches. 
For the case when we ignore tSZ-noise from all haloes (no-tSZ), we set the second term \mbox{$\mathbf{\hat{C}}_{\rm tSZ} = 0$}. 
We calculate \mbox{$\mathbf{\hat{C}}_{\rm tSZ} \equiv \mathbf{\hat{C}}_{\rm tSZ}^{\rm Gaussian}$} using Gaussian realizations of the tSZ signals for the other two cases. 
For Poisson-tSZ, since the noise from tSZ will be lower (higher) in pixels where massive clusters are absent (present), we apply a scaling factor to \mbox{$\mathbf{\hat{C}}_{\rm tSZ}$} during the likelihood calculation. 
However, we do not find the results to be sensitive to the choice of this scaling factor. 
This simple scaling does not treat the effect of tSZ non-Gaussianities but we do not notice any bias in our results and hence consider them to be negligible. 
We provide more details about the scaling in Appendix~\ref{appendix_snr_calculation}.

We perform \howmanysimulations{} simulations and use the distribution of best-fits  $\recoveredYsymbolbold_{\rm fit}$ to compute the $\snr$.
Specifically, $\snr$ is calculated as the inverse of the $1\sigma$ uncertainty defined using the $16^{\rm th}$ and the $84^{\rm th}$ percentiles. 
We do not use the standard deviation of the distributions to estimate the $\snr$ as that will be dominated the non-Gaussian tails for the Poisson-tSZ case. 
See Appendix~\ref{appendix_snr_calculation} for more details.
Using this setup, we construct an experiment-dependent $\snr$ look-up table for clusters in a ($\mvir, z$) grid and use that to estimate the limiting cluster mass that satisfies the detection threshold $\snr \ge \snrlimit$ as a function of redshift for each experiment. 
The ($\mvir, z$) grid we chose is:
\mbox{$\Delta {\rm log}(\mvir/\msol) \in [13,15.4]$} with \mbox{$\Delta{\rm log} (M/\msol) = 0.1$} and $z \in [0.1, 3]$ with $\Delta z = 0.1$.

\begin{figure*}
\centering
\ifdefined\ApJsubmit
\includegraphics[width=\textwidth, keepaspectratio]{limiting_masses_vs_z_stageallexps_nomasking.pdf}
\else
\includegraphics[width=\textwidth, keepaspectratio]{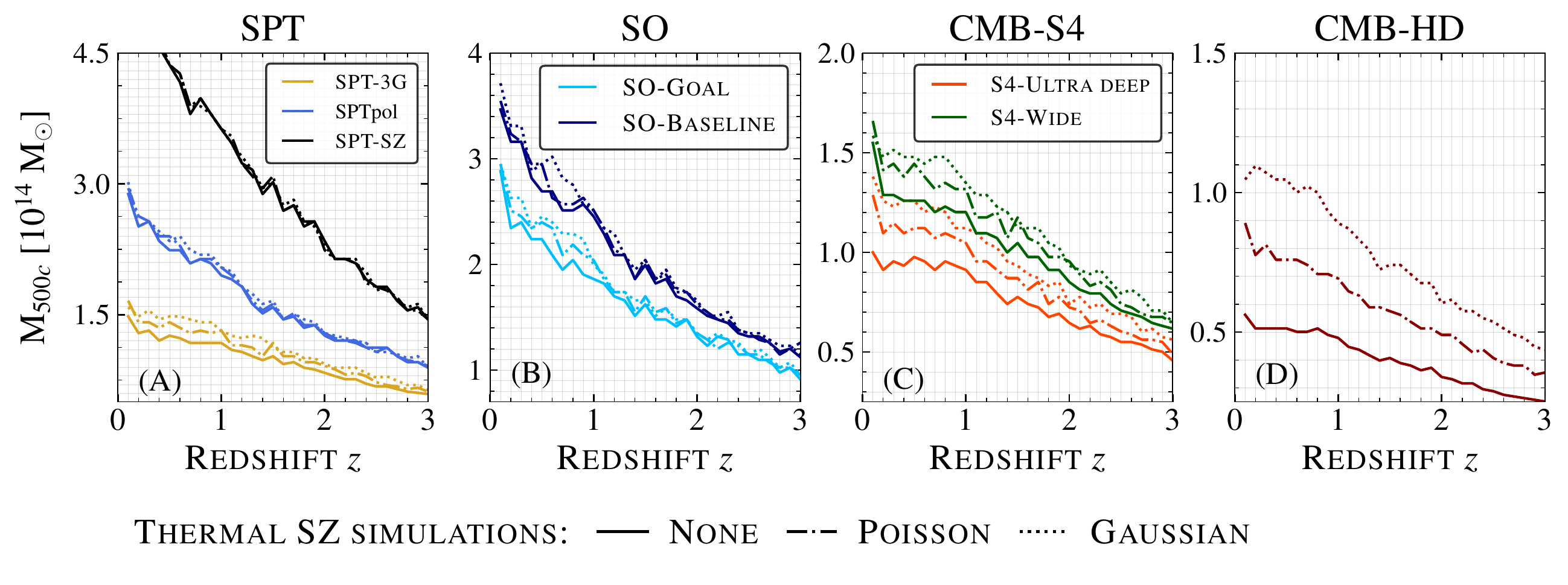}
\fi
\caption{
Cluster survey sensitivity expressed using the minimum detectable cluster mass at each redshift. 
For each experiment, the three curves correspond to different simulations of the tSZ signals: solid for no, dash-dotted for Poisson, and dotted for Gaussian-tSZ cases. 
For \sptsz, \sptpol, \sofid{} and \sogoal, all the three curves lie on top of each other indicating that the tSZ-noise is not important. 
For \sptthreeg{} (yellow in panel A) and \sfour{} (green in panel C), although the sensitivity for the Gaussian-tSZ is worse than the other two, the difference between the three curves is small $\le 10\%$. 
For \cmbhd{} (\sfourdeep) shown as darkred in panel D (orange in panel 3), the Poisson case reduces the sensitivity by $35-50\%$ ($10-15\%$) compared to the no-tSZ case. 
Switching to Gaussian-tSZ simulations can degrade the sensitivity further for these two surveys. 
}.
\label{fig_limiting_mass_vs_z_all_exps}
\end{figure*}

\section{Fisher formalism}
\label{sec_fisher_formalism}

In this section, we describe the Fisher forecasting formalism to combine cluster counts with primary CMB TT/EE/TE spectra. 
We also check the gain by including information from the tSZ power spectrum although the constraining power of $\binnedcluscounts$ is higher than the tSZ spectrum since we are localizing the signal better in numerous mass, redshift, and $\snr$ bins.
The procedure is briefly described below and we refer the reader to \citetalias{raghunathan21} for more details. 

To obtain the binned cluster counts $\binnedcluscounts$, we use the Monte Carlo (MC) sampling approach described in \citetalias{raghunathan21}. 
We choose 40 and 25 logarithmic bins for lensing mass and tSZ $\snr$: $M_{L} \in  [10^{12}, 10^{16}]\ \msol$ and $q \in[5, 500]$.
For redshift, 
we use $\Delta z = 0.1$ for $0.1 \le z < 1.5$ and group all $z\ge1.5$ clusters in one massive redshift bin. 
To this end, we obtain the number of haloes $n(\mvir,z)$ in the range $\mvir \in  [10^{13}, 10^{16}]\ \msol$ with $\Delta {\rm} \mvir = 10^{12}\ \msol$ and $0.1 \le z \le 3$ with $\Delta z = 0.1$ using \citet{tinker08} HMF. 
We assign a tSZ flux and an associated tSZ $\snr$ $q$ for each halo using the distributions: $\mathcal{N}\left( {\rm log}\Ysz | M, z, \sigma_{{\rm log}_{\Ysz}} \right)$ and $\mathcal{N}\left( q | \Ysz/\sigma_{\Ysz}, 1\right)$. 
For tSZ $\snr$, we use the $\snr$ look-up table described in \S\ref{sec_cluster_detection}.
Similarly, lensing mass and redshifts are assigned using the distributions: $\mathcal{N}\left( M_{L} |M, \sigma_{M_{L}} \right)$ and $\mathcal{N}\left( z |z_{\rm true}, \sigma_{\rm z} \right)$ where $\sigma_{M_{L}}$ is the CMB-cluster lensing mass error calculated using both CMB temperature and polarization measurements for each survey \citep{raghunathan17a, raghunathan19b}; and the redshift errors are assumed to be $\sigma_{z} = 0$. 
External lensing information, like for example from optical surveys, is ignored for simplicity. 
The derivatives $\partial \binnedcluscounts/\partial \theta$ as a function of parameter $\theta$ for Fisher forecasts were obtained with the finite difference approach using the same setup above and the weighting schemes described in \citetalias{raghunathan21}. 

For primary CMB TT/EE/TE Fisher matrix, we use the lensed CMB spectra obtained using \texttt{CAMB} \citep{lewis00} software for the fiducial \planck{} 2015 cosmology. 
Data from all bands in a given survey are optimally combined using the ILC approach. 
The assumptions about polarized foregrounds are similar to \citetalias{raghunathan21}: 2\% (3\%) polarization fractions for DSFG (radio galaxies) consistent with measurements from ACT \citep{datta18} and SPT \citep{gupta19} and the diffuse kSZ/tSZ signals are assumed to be unpolarized. 
We use information up to $\ell_{\rm max} = 3500$ for the current and $\ell_{\rm max} = 5000$ for the future CMB surveys. 

In this work, we also 
assess the improvement gained by including information from the tSZ power spectrum $\clyy$. 
We obtain the $\clyy$ and the derivatives
$\partial \clyy/\partial \theta$ using the Poisson simulations described in \S\ref{sec_diffusetsz}. 
Although \citet{hurier17} argue that the correlation between $\binnedcluscounts$ and $\clyy$ is minimal, we follow two approaches. 
In the first approach, we use the tSZ power spectrum in Fig.~\ref{fig_poisson_tsz_simulation_power_spectra} obtained without masking any detected clusters. 
The second approach is highly conservative where we compute the $\clyy$ after masking clusters detected by the survey under consideration. 
Masking the detected clusters should significantly reduces the correlation between the two probes. 
We set $\ell_{\rm max} = 8000$ for $\clyy$.  

The parameter $\theta$ constrained in this work is one of the 15 observable-mass scaling relation or cosmological parameters. 
The parameters governing the observable-mass scaling relation are given in Eq.(\ref{eq_Ysz_mass}) and Eq.(\ref{eq_Ysz_mass_scatter}): \mbox{$\theta \in [\alphay, \hsebias, \betay, \gammay, 
\sigma_{{\rm log}Y,0}, \alphasigmalogy, \gammasigmalogy]$}. 
The cosmological parameters are the 6 \lcdm{}  parameters along with the extension to include the sum of neutrino masses and dark energy equation of state: \mbox{$\theta \in [A_{s},\ h,\ \summnu,\ n_{s},\ \omchsq,\ \ombhsq,\ \taure,\ \wde]$}.
Note that we fix the normalization $Y_{\ast}$ since it is highly degenerate with the hydrostatic mass bias parameter $1-\hsebias$. Including $Y_{\ast}$, however, has a negligible impact on other scaling relation or cosmological parameter constraints.

\section{Results and discussion}
\label{sec_results}

\subsection{Cluster detection sensitivity}
\label{sec_baseline_results}
In Fig.~\ref{fig_limiting_mass_vs_z_all_exps} we present the minimum detectable cluster mass satisfying $\snr \ge \snrlimit$ at each redshift $z \in [0.1, 3.0]$ in bins of $\Delta z = 0.1$ for all experiments ordered from left to right based on the survey timeline. 
Each experiment consists of three curves: solid for no-tSZ, dash-dotted for Poisson-tSZ, and dotted for Gaussian-tSZ cases. 
The difference between no (solid) and Poisson (dash-dotted) cases quantifies the impact of the tSZ-noise. 
The difference between the Poisson (dash-dotted) and the Gaussian (dotted) cases shows the importance of the non-Gaussian distribution of the tSZ but this is only important when the tSZ-noise is non-negligible. 
This can be inferred from the figure where we can note that although 
the $\snr$ is worse for the Gaussian case compared to the Poisson, 
the level of $\snr$ degradation depends on the experiment. 
As expected, the results are better in the absence of the tSZ signals (solid curves) compared to the other two cases.

For \cmbhd, assuming a Gaussian distribution of the diffuse tSZ 
increases the limiting mass at all redshifts by $\sim 25-40\%$ compared to the Poisson case. 
Both cases are worse than the no-diffuse tSZ case: Poisson by 40-50\% and Gaussian by $\ge80\%$. 
Thus, we find that the tSZ-noise can degrade the cluster $\snr$ significantly for the \cmbhd{} experiment.
For \sfourdeep{}, the $\snr$ reduction for the Gaussian case compared to Poisson case is around $15\%$. 
Compared to the no-tSZ case, the Poisson case is worse by $10-15\%$, indicating that the tSZ-noise is mildly important for \sfourdeep.
For \sfour{} and \sptthreeg{} surveys, the difference between Gaussian and Poisson case is much smaller $\sim 5\%$ and the Poisson case is worse than the no-diffuse tSZ case by $\lesssim 8\%$. 
In the case of other experiments namely \sogoal, \sofid, \sptpol{} or \sptsz{}, the three curves are indistinguishable and the tSZ-noise has no impact on these surveys. 

The reason that the tSZ-noise has a large impact only on \cmbhd{} is primarily because of the reduced level of residual CIB signals in the Compton-$y$ maps. 
The residual CIB is non-negligible for \sfourdeep{} and hence the impact of tSZ confusion noise is lower than \cmbhd. 
Both \sfour{} and \sptthreeg{}, despite having low instrumental noise compared to current surveys, are limited by the residual CIB signals. 
The sensitivity of other experiments are limited by both the instrumental noise and the residual foreground signals. 
The importance of the noise from tSZ signals reduces when the residual noise or foreground signals increase.
This is also evident from Fig.~\ref{fig_ilc_residuals} where we can note that the total residual noise is roughly an order of magnitude lower than the signal (gray band) at arcminute scales for \cmbhd{} while that is not the case for other experiments. 

\subsection{Results after removing the detected clusters}

\begin{figure*}
\centering
\ifdefined\ApJsubmit
\includegraphics[width=0.7\textwidth, keepaspectratio]{compton_y_powerspec_all_exps.pdf}
\else
\includegraphics[width=0.7\textwidth, keepaspectratio]{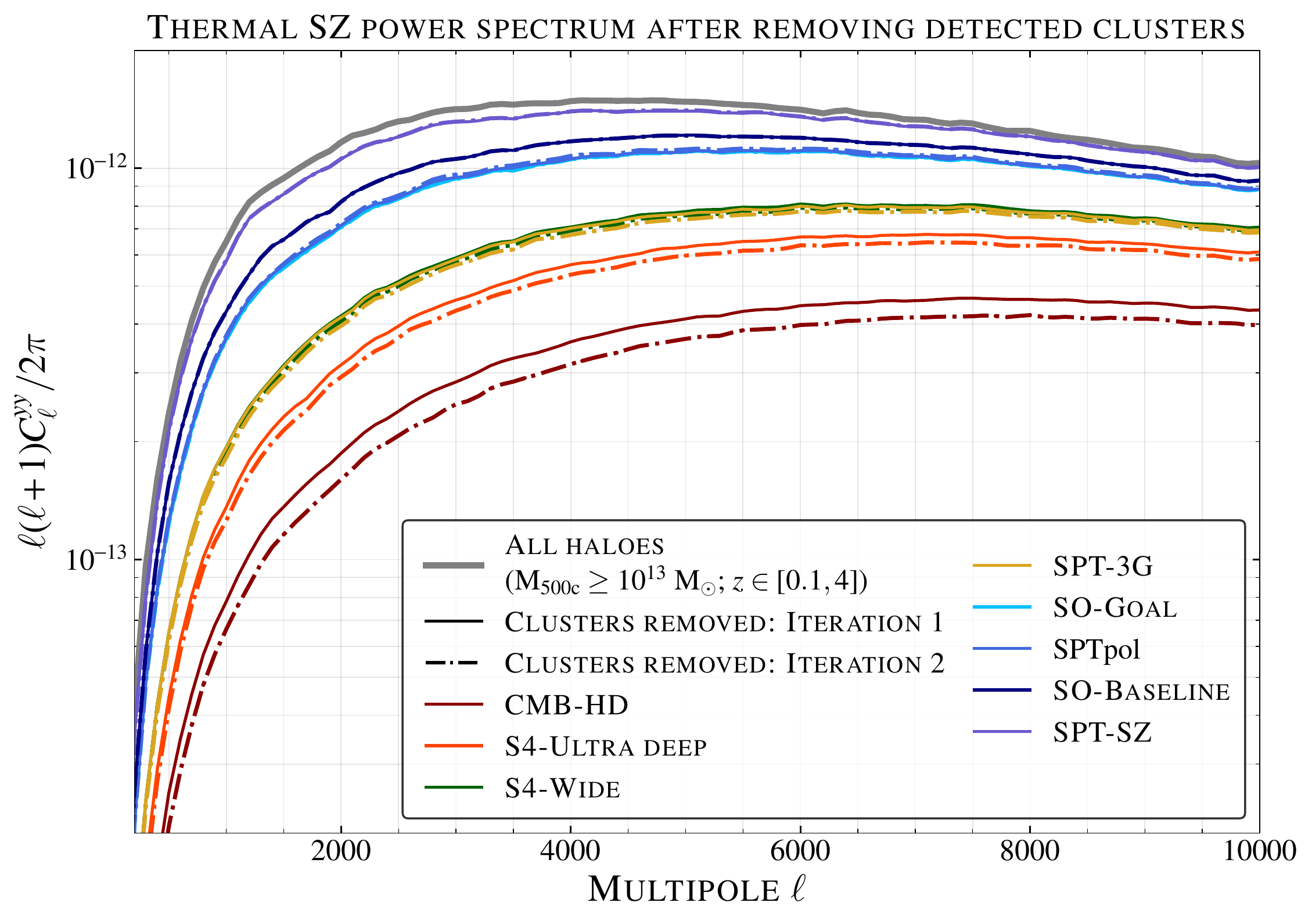}
\fi
\caption{Impact of masking the detected clusters on the Compton-$y$ power spectrum are shown as colored curves. 
For reference, the fiducial unmasked spectrum is in grey.
After masking we note a suppression of the Compton-$y$ power, measured at $\ellnorm$, for different experiments in the following order: \cmbhd{} ($\szpowerreduccmbhd$); \sfourdeep{} ($\szpowerreducsfourdeep$); \sfour{} and \sptthreeg{} ($\szpowerreducsfoursptthreeg$); \sogoal{} and \sptpol{} ($\szpowerreducsogoalsptpol$); \sofid{} ($\szpowerreducsofid$); and \sptsz{} ($\szpowerreducsptsz$).
The solid and dash-dotted colored curves are for two levels of masking. 
The second level of masking is only important for experiments where the tSZ-noise is important namely \cmbhd{} and \sfourdeep. 
After second iteration, we see a further 13\% (6\%) suppression in the power for \cmbhd{} (\sfourdeep).
All curves correspond to the median value from \howmanysimulations{} Poisson realizations.
}
\label{fig_compton_y_power_spec_all_exps}
\end{figure*}

\begin{figure}[b]
\centering
\ifdefined\ApJsubmit
\includegraphics[width=0.49\textwidth, keepaspectratio]{poisson_sims_zoomed.pdf}
\else
\includegraphics[width=0.49\textwidth, keepaspectratio]{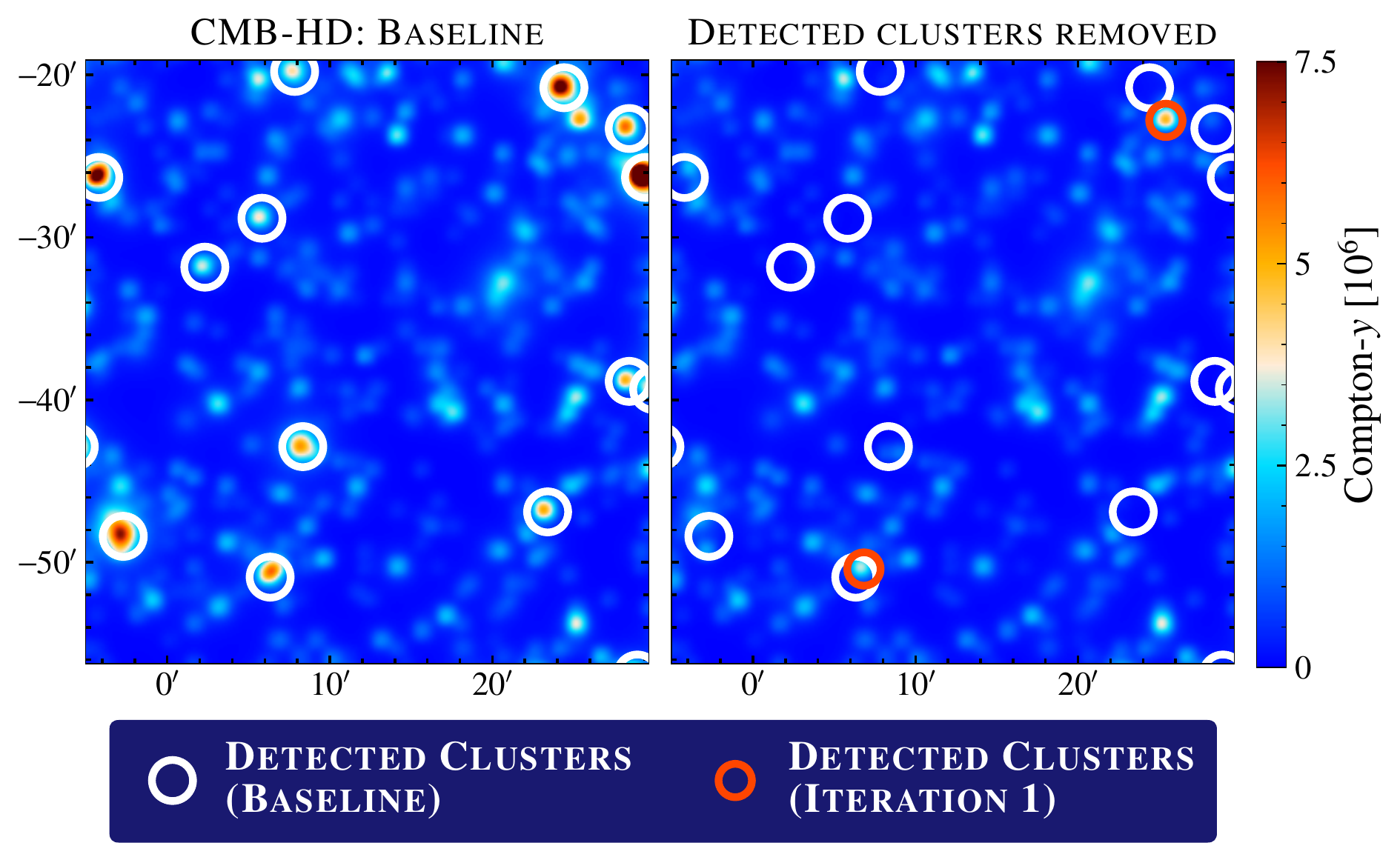}
\fi
\caption{
Zoomed-in version of the Poisson realization shown in Fig.~\ref{fig_poisson_tsz_simulation}. 
The region shown is $35^{\prime} \times 35^{\prime}$ wide and centered at $(\theta_{x}, \theta_{y})= (17.^{\prime}3, 18.^{\prime}5)$. 
{\it Left panel:} Clusters detected above $\snr \ge \snrlimit$ by the \cmbhd{} experiment are highlighted using white circles. 
{\it Right panel:} Removing the clusters detected in the baseline case (left panel) allows us to detect new smaller (less massive or distant) clusters near the same LOS. 
The new detections are highlighted using red circles.
}
\label{fig_poisson_tsz_simulation_after_masking}
\end{figure}

Now we turn to the effect of removing the detected clusters from the maps. 
\subsubsection{Reduction in tSZ power after masking}
\label{sec_tsz_power_after_masking}
Given that the detected clusters dominate the tSZ power spectrum and their number counts is highly non-Gaussian, 
masking the detected clusters should have an impact on the tSZ power spectrum. 
In Fig.~\ref{fig_compton_y_power_spec_all_exps}, we present the resultant Compton-$y$ power spectrum after masking clusters detected with $\snr \ge \snrlimit$ by each experiment. 
The curves represent the median value from \howmanysimulations{} Poisson realizations.
The power spectrum for the baseline case, without any masking, is shown as the thick grey curve. 
The masked power spectra are shown as solid curves in different shades for each experiment. The dash-dotted curves are for a second level of masking and they are discussed next. 
We use the value at $\ell_{\rm norm} = 3000$ to quote the reduction in Compton-$y$ power after cluster masking. 
For \sptsz{}, masking the detected clusters does not have a huge impact on the power spectrum and only reduces it by $\szpowerreducsptsz$. 
For \sofid, we find that masking the detected clusters suppresses the power by $\szpowerreducsofid$. 
The reduction in power is roughly similar for \sptpol{} ($\szpowerreducsptpol$) and \sogoal{} ($\szpowerreducsogoal$) followed by \sptthreeg{} and \sfour{} where the power goes down by more than a factor of two ($\szpowerreducsfoursptthreeg$). 
For \sfourdeep{} and \cmbhd{}, we find an even higher suppression: $\szpowerreducsfourdeep$ and $\szpowerreduccmbhd$ respectively. 
Switching from median to mean value of the \howmanysimulations{} Poisson realizations, decreases the power further by $\sim10\%$ for all experiments. 
This reduction in the Compton-$y$ power and the performance of different experiments are fully in accordance with the inference from Fig.~\ref{fig_limiting_mass_vs_z_all_exps}. 

Next, we focus on the colored dash-dotted curves in the Fig.~\ref{fig_compton_y_power_spec_all_exps} which are for the next level of cluster masking 
(i.e:) after removing the first set of detected clusters, we re-run the cluster $\snr$ computation to detect new clusters (see \S\ref{sec_survey_sensitivity_after_masking} below). 
For \cmbhd{} (\sfourdeep), a second round of masking lowers the Compton-$y$ power further down by 13\% (6\%).
The solid and dash-dotted lines are almost indistinguishable in Fig.~\ref{fig_compton_y_power_spec_all_exps} for other experiments, since the tSZ-noise is negligible and removing detected clusters does not help in finding new ones. 

\subsubsection{Survey sensitivity after removing detected clusters and the tSZ confusion noise floor}
\label{sec_survey_sensitivity_after_masking}

\begin{figure}
\centering
\ifdefined\ApJsubmit
\includegraphics[width=0.49\textwidth,  keepaspectratio]{limiting_masses_vs_z_stageallexps_detected_clusters_removed.pdf}
\else
\includegraphics[width=0.49\textwidth,  keepaspectratio]{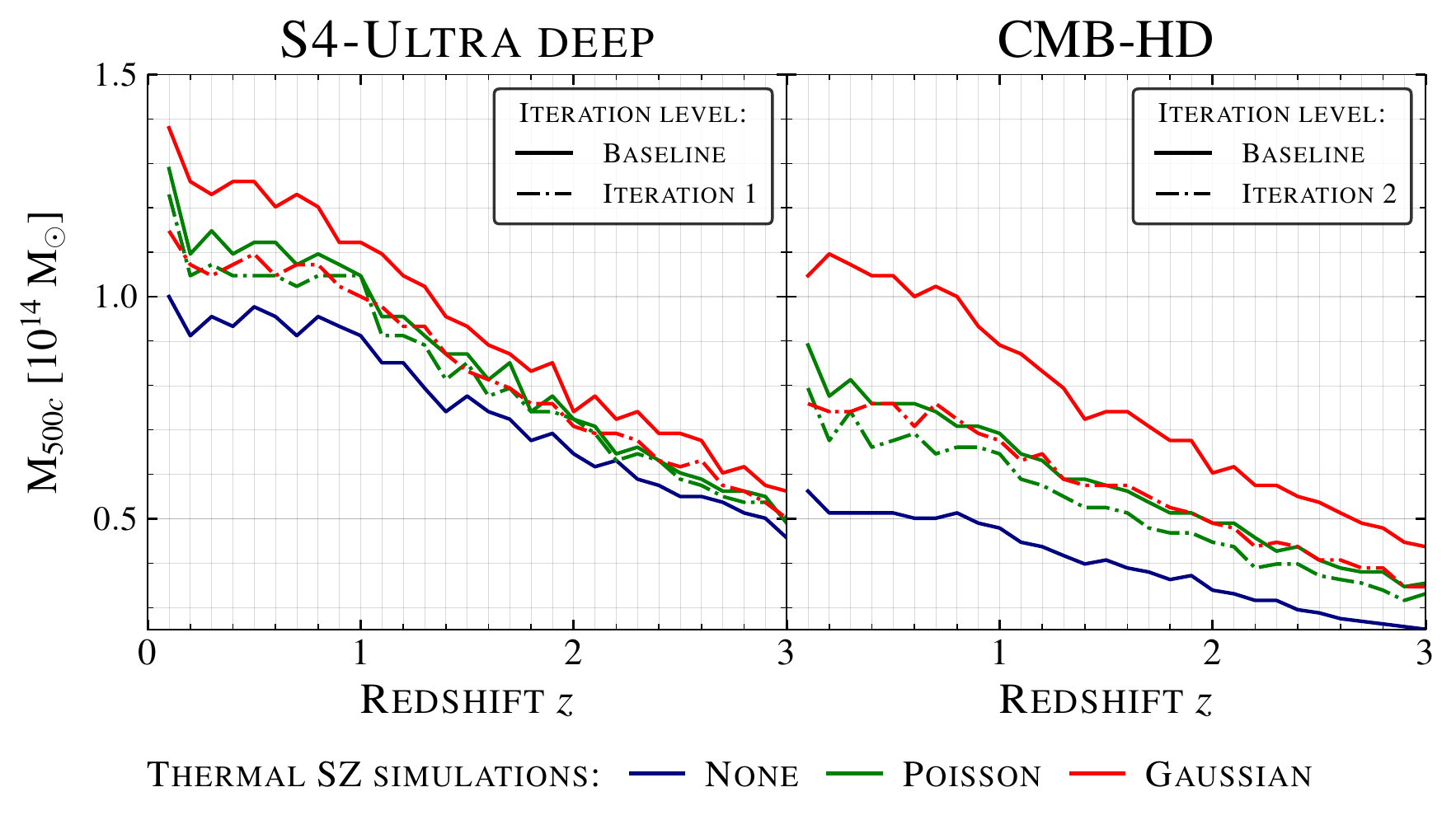}
\fi
\caption{Change in the cluster survey sensitivity after removing the detected clusters for \sfourdeep{} (\cmbhd) in the {\it left} ({\it right}) panel. The three colors are for different simulations of the tSZ signals: blue for no, green for Poisson, and red for Gaussian. 
Solid and dash-dotted curves represent results from the baseline setup vs the ones after removing detected clusters. 
After one round of cluster removal, Poisson and Gaussian cases match for \sfourdeep. 
On the other hand, \cmbhd{} requires multiple iterations for the Poisson and Gaussian cases to roughly match (within $\lesssim 8\%$). 
Even after removing detected clusters, the Poisson-tSZ is higher than the no-tSZ case by $\cmbhdtsznoisefloor$ ($\sfourdeeptsznoisefloor$) for \cmbhd{} (\sfourdeep) which represents the systematic floor set by the confusion noise from haloes below the detection limit.
}
\label{fig_limiting_mass_vs_z_after_masking}
\end{figure}

The process of removing detected clusters facilitates us to detect new clusters. 
This is illustrated in Fig.~\ref{fig_poisson_tsz_simulation_after_masking} for the \cmbhd{} experiment. 
We assume that detected clusters can be perfectly modeled and removed from the maps using a template fitting approach. 
This, however, is not true in reality and we will have residuals due to the mis-match between the assumed templates and the true cluster signal arising due to mergers or asymmetric cluster profiles. 
We ignore such complexities since we are only trying to demonstrate the proof of concept in this work. 
The new detections are less massive or distant clusters which were missed in the first iteration because they lie near the LOS of a more massive cluster. 
In the figure, the baseline case is shown in the {\it left} panel and the locations of clusters detected by the \cmbhd{} experiment are highlighted as white circles. 
The second run after perfect cluster removal is shown in the {\it right} panel and the newly detected clusters are highlighted using red circles. 
The highlighted object in the bottom left is a cluster with $(\mvir, z) = (7 \times \Mminredefine\ \msol, 0.7)$ and is detected in the second run after removing a $(\mvir, z) = (9 \times \Mminredefine\ \msol, 0.8)$ in the same LOS. 
The cluster detected in the top right is superposition of two clusters: $\mvir \sim 6 \times \Mminredefine\ \msol$ at $z = 1.2$ which is just below the detection threshold and another cluster $\mvir \sim 3 \times \Mminredefine\ \msol$ at $z = 1.8$. 
Since removing the detected clusters improves the overall survey sensitivity slightly (see Fig.~\ref{fig_limiting_mass_vs_z_after_masking}), the former is detected in the second run.
This figure is the zoomed-in version ($35^{\prime} \times 35^{\prime}$) centered at $(\theta_{x}, \theta_{y})= (17.^{\prime}3, 18.^{\prime}5)$ of the Poisson realization shown in Fig.~\ref{fig_poisson_tsz_simulation}. 

Since removing the detected clusters reduces the tSZ power spectrum as shown in Fig.~\ref{fig_compton_y_power_spec_all_exps}, the cluster sensitivity also improves. 
We quantify this improvement now and also check if multiple iterations of removing the detected clusters can help in completely eliminating the impact of noise from tSZ signals. 
We limit these calculations to \sfourdeep{} and \cmbhd{} since the tSZ-noise does not have significant impact on the other surveys. 
The results are presented in Fig.~\ref{fig_limiting_mass_vs_z_after_masking} with \sfourdeep{} in {\it left} and \cmbhd{} in {\it right} panels respectively. 
Solid lines are for the baseline case (same as Fig.~\ref{fig_limiting_mass_vs_z_all_exps}) and dash-dotted lines are after removing the detected clusters. 
We compare the results from Poisson (green) and Gaussian-tSZ (red) cases for the successive iterations. 
The no-tSZ (blue) case is also shown in the figure for reference. 
The underlying power spectrum used for Gaussian realizations of the tSZ after removing detected clusters are the colored lines in Fig.~\ref{fig_compton_y_power_spec_all_exps} for the respective experiment.

For \sfourdeep{}, after the first iteration of detected-cluster removal, the Poisson and Gaussian cases roughly overlap. 
They are both $\sim \sfourdeeptsznoisefloor$ above the blue (no-tSZ case) and we find no further improvement in the successive iterations. 
This is because the tSZ noise is now sourced from the haloes below the detection limit for \sfourdeep.
The results are slightly different for \cmbhd{} on the other hand which requires two levels of cluster removal for the Poisson and Gaussian cases to match to within $\lesssim 8\%$.  
However, even after two rounds of masking, the green and the red curves are higher than the blue (no-tSZ) by $\cmbhdtsznoisefloor$. 
The above results suggest that the tSZ confusion noise from haloes below the detection limit sets an ultimate floor of $\sim \cmbhdtsznoisefloor{} (\sfourdeeptsznoisefloor)$ for cluster detection using \cmbhd{} (\sfourdeep). 
There is also a small improvement in sensitivity, ratio of green solid to dash-dotted lines, for both experiments after masking: \cmbhdsensitivityimprovement{} (\sfourdeepsensitivityimprovement) for \cmbhd{} (\sfourdeep). 


\subsubsection{Cluster counts and parameter constraints}
\label{sec_parameter_constraints}

\begin{deluxetable}{| l | c | c | c | c | c | c |}
\def\arraystretch{1.2}
\tablecaption{Expected number of $\snr \ge \snrlimit$ clusters from SZ surveys.}
\label{tab_cluster_counts}
\tablehead{
\multirow{2}{*}{Experiment} & \multicolumn{3}{c|}{Total clusters} &  \multirow{2}{*}{$z^{\rm med}$} & $M^{\rm med}_{500c}$
\\
\cline{2-4}
& Total & $z \ge 1.5$ & $z \ge 2$ & & [$10^{14}\ \msol]$
}
\startdata
\sptsz & 410 & 7 & - & 0.6 & 3.6  \\\hline
\sptpol & 600 & 24 & 3 & 0.7 & 2.5 \\\hline
\sptthreeg & 6935 & 477 & 80 & 0.7 & 1.3 \\\hline\hline
\sofid & 14424 & 490 & 53 & 0.7 & 2.5 \\\hline
\sogoal & 26445 & 1256 & 189 & 0.7 & 2.0 \\\hline\hline
\sfour & 107747 & 7958 & 1548 & 0.8 & 1.6 \\\hline
\sfourdeep & 11801 & 1144 & 231 & 0.8 & 1.0 \\\hline\hline
\cmbhd & 514530 & 79099 & 20682 & 0.9 & 0.6 \\\hline\hline
\enddata
\end{deluxetable}

\begin{figure*}
\centering
\ifdefined\ApJsubmit
\includegraphics[width=0.9\textwidth, keepaspectratio]{parameter_constraints_clusters_cmb.pdf}
\else
\includegraphics[width=0.9\textwidth, keepaspectratio]{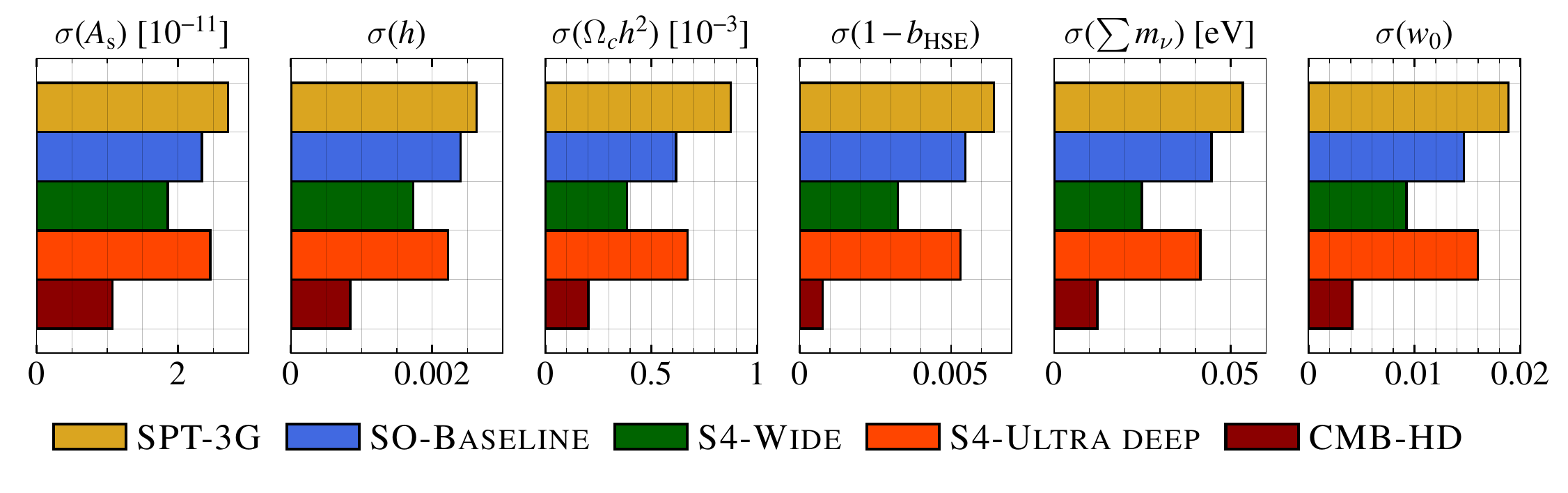}
\fi
\caption{Marginalized parameter constraints for the current (\sptthreeg) and future (\sofid, \sfour, \sfourdeep{} and \cmbhd) surveys obtained by combining cluster abundance measurements and the lensed CMB power spectra (TT/EE/TE). 
A \planck-like prior $\sigma(\taure) = 0.007$ has been assumed in all cases. 
We find that the all surveys can obtain $\lesssim 1\%$ constraints on $\sigma(A_{s}),\ \sigma(h),\  \sigma(\omchsq)$ and $\sigma(1-\hsebias)$. 
Furthermore, the errors on $\sigma(\wde)$ can be reduced to $\le 2\%$ in the next few years by \sptthreeg{} and \sofid.
\cmbsfour{} and \cmbhd{} can strengthen these measurements further by reducing $\sigma(\wde)$ to sub-percent levels and by enabling a $3-5\sigma$ detection of the sum of neutrino masses.
}
\label{fig_parameter_constraints}
\end{figure*}

In Table~\ref{tab_cluster_counts}, we show the expected number of clusters along with the median mass $\mvir$ and redshift. 
We also explicitly show the number of high redshift, $z \ge 1.5$ and $z \ge 2$, clusters. 
For \sfour, we ignore the regions contaminated by the galactic foregrounds and use $\fsky=50\%$. 
The sky fractions for other surveys are given in Table~\ref{tab_exp_specs}. 
These calculations use the mass thresholds derived using the Poisson tSZ simulations (dash-dotted curves in Fig.\ref{fig_limiting_mass_vs_z_all_exps}). 
Switching to the Gaussian-tSZ case reduces the number of clusters for \cmbhd{} (\sfourdeep) by $\times 1.8\ (1.3)$.
On the other hand, ignoring noise due to tSZ signals from other haloes increases the total clusters by $\times 2.3$ and  $\times 1.4$ for \cmbhd{} and \sfourdeep{} respectively. 
The impact due to the choice of the tSZ signals is smaller for all the other surveys which is consistent with the mass thresholds reported above.
The expected number of clusters, the median mass $M^{\rm med}_{500c}$, and the median redshift $z^{\rm med}$ obtained in this work are in reasonable agreement with other works in the literature, for example, the published SPT-SZ catalog by \citet{bleem15} and the forecasts presented by \citet{SO18} for the two SO configurations.

We use the cluster counts and combine them with the primary CMB TT/EE/TE spectra to obtain parameter constraints. 
The results with the inclusion of tSZ power spectrum is discussed next.
If Fig.~\ref{fig_parameter_constraints}, we show the marginalized parameter constraints for $\theta \in [A_{s},\ h,\ \omchsq, \summnu, \wde]$.
We limit these to the current (\sptthreeg) and future (\sofid, \sfour, \sfourdeep{} and \cmbhd) surveys. 
We find that all the surveys can reduce the uncertainties on $A_{s},\ h,\ \omchsq{}$ and the hydrostatic mass bias $1-\hsebias$ to sub-percent levels. 
In addition, \sptthreeg{} and \sofid{} can reduce $\sigma(\wde)$ to $\lesssim 2\%$. 
On the other hand, \cmbsfour{} and \cmbhd{} can obtain sub-percent level constraints on $\sigma(\wde)$. 
Assuming a normal hierarchy, both the surveys can also enable a $3-5\sigma$ detection of the sum of neutrinos masses.
The constraints, $\sigma(\summnu)$ in particular, depends on the choice of prior used for $\sigma(\taure)$. 
Swapping the baseline \planck-like prior $\sigma(\taure) = 0.007$ to LiteBIRD-like prior $\sigma(\taure) = 0.002$ (no prior) improves (degrades) $\sigma(\summnu)$ by $\times1.3-1.5$. 

When the noise from tSZ signals are fully ignored, the \cmbhd{} (\sfourdeep) constraints improve only by $10-15\%$ ($5-10\%$) for all parameters. 
Although the number of clusters go up by significantly in this case as mentioned above, their impact on parameter constraints in only marginal. 
This is because the constraints are driven by the different parameter degeneracy directions between clusters vs CMB and including less massive haloes by lowering the mass thresholds does not improve the constraining power significantly. 

Now, we present the improvement in the constraining power with the addition of the tSZ power spectrum. 
We check the results with both the unmasked $\clyy$ (black curve in Fig.~\ref{fig_poisson_tsz_simulation_power_spectra}) and $\clyy$ after masking detected clusters in each survey (colored curves in Fig.~\ref{fig_compton_y_power_spec_all_exps}). 
Note that masking the detected clusters should significantly reduce the correlation between cluster counts and $\clyy$. 
As mentioned in \S\ref{sec_fisher_formalism}, we use $\ell_{\rm max} = 8000$ for $\clyy$. 
While small-scale residual CIB signals can be potentially problematic and need to be carefully modeled, we ignore CIB modeling given that the improvement after adding $\clyy$ is not dramatic. 
The maximum constraining power is for $\sigma(\omchsq)$ which improves by $20-25\%$ ($8-10\%$) for all surveys with the unmasked (masked) $\clyy$. 
For other parameters, the improvement is $\lesssim10\%$ even in the optimistic case of using the unmasked $\clyy$. 

\section{Conclusion}
\label{sec_conclusion}

We assessed the importance of noise from tSZ signals, arising from the haloes along the LOS and the ones below the detection limit, for cluster detection from a variety of CMB surveys ranging from the past to the future namely: \sptsz, \sptpol, \sptthreeg, \sofid, \sogoal, \sfour, \sfourdeep{} and \cmbhd. 
The framework involved three different ways of injecting the tSZ signals into the simulated maps which contain CMB, experimental noise, and other astrophysical foregrounds. 
The tSZ simulations includes contribution from all haloes in the mass range $\mvir \ge \Mminredefine\ \msol$ at redshifts $z \in [\zmin, \zmax]$ and are simulated either as Poisson or as Gaussian realizations. 
For Poisson realizations, we adopt \citet{tinker08} HMF to get the halo number counts and model the cluster tSZ signal using the generalized NFW profile. 
The power spectrum of the Poisson realizations was used to generate Gaussian realizations. 
We compared these results with the ones where the tSZ signals (no-tSZ) are fully ignored. 
The difference between Poisson and the no-tSZ cases was used to quantify the importance of the noise from tSZ signals. 

Our results indicated that the tSZ-noise has a significant impact ($35-50\%$) only on the \cmbhd{} experiment and mildly ($10-15\%$) affects CMB-S4's delensing (\sfourdeep) survey. 
For all the other experiments, the tSZ signals either have no impact (\sptsz, \sptpol, \sofid{} and \sogoal) or lead to minor (\sptthreeg{} and \sfour) differences in the results compared to the no-tSZ case. 
This is because the Compton-$y$ maps of these experiments are dominated by residual experimental noise or other astrophysical foregrounds which are much higher than fiducial tSZ signal level.

In all cases, we found that the Poisson-tSZ case returns a higher $\snr$ than the Gaussian-tSZ case for a given cluster. 
This, however, is only important when the tSZ-noise has a large impact in the survey sensitivity. 
In other words, performing a Gaussian realization of the tSZ signals: (a) does not have any impact on \sptsz, \sptpol, \sofid{} and \sogoal{}; (b) can slightly ($5-10\%$) worsen the results for \sptthreeg{} and \sfour; and (c) can significantly affect \sfourdeep{}  ($15-20\%$) and \cmbhd{} ($\ge 80\%$). 

We also quantified the impact of masking the detected clusters on the tSZ power spectrum for all the surveys.
The reduction in the power is quoted at $\ellnorm$. 
Compared to the original (unmasked) case, the power spectrum goes down by: $\szpowerreduccmbhd$ for \cmbhd; $\szpowerreducsfourdeep$ of \sfourdeep; $\szpowerreducsfoursptthreeg$ for \sfour{} and \sptthreeg; $\szpowerreducsogoalsptpol$ for \sogoal{} and \sptpol; $\szpowerreducsofid$ for \sofid; and $\szpowerreducsptsz$ for \sptsz. 
This reduction can have important implications for detecting the kSZ power spectrum in the small-scale CMB TT power spectrum for current and future CMB surveys since the amplitude of the kSZ is degenerate with that of the tSZ \citep[see for example,][]{dunkley13, george15, reichardt21}. 

We also showed that a perfect removal of the detected clusters can help us detect less massive and distant clusters that lie near the LOS of the removed clusters. 
Removing the detected clusters also improved the cluster survey sensitivity slightly by $\sim$\cmbhdsensitivityimprovement{} (\sfourdeepsensitivityimprovement) for \cmbhd{} (\sfourdeep). 
The improvement saturates after two (one) round of cluster removal and the tSZ confusion noise, sourced by haloes below the detection limit, sets a floor of \cmbhdtsznoisefloor{} (\sfourdeeptsznoisefloor) for cluster detection using \cmbhd{} (\sfourdeep) compared to the case when tSZ signals are fully ignored. 

We forecasted the expected number of clusters from all the surveys finding significant improvement in the sample size compared to the currently available SZ catalogs. 
Finally, we combined the cluster abundance measurements with primary CMB TT/EE/TE and tSZ power spectra to obtain constraints on cluster observable-mass scaling relation and cosmological parameters for current and future surveys. 
In the next few years, \sptthreeg{} and \simonsobs{} will be able to constrain the error on the dark energy equation of state parameter $\sigma(\wde)$ to an accuracy level of $\le 2\%$. 
The \cmbsfour{} experiment, in the next decade, can improve the accuracy on $\sigma(\wde)$ to sub-percent levels and also enable a $3\sigma$ detection of the sum of neutrino masses $\summnu$. 
The futuristic \cmbhd{} survey will add further constraining power and is also capable of detecting $\summnu$ at $\ge 5\sigma$. 
The simulation products and results are publicly available and can be downloaded from this \href{https://github.com/sriniraghunathan/tSZ_cluster_forecasts}{link$^{\text{\faGithub}}$}.

\section*{Acknowledgments}
I am indebted to Gil Holder and Nathan Whitehorn for their feedback and suggestions throughout the course of this work. 
I also thank Tom Crawford, Daisuke Nagai, Yuuki Omori and Christian Reichardt for useful discussions and their feedback on the manuscript. 
Finally, I thank the anonymous referee for useful suggestions that helped in shaping this manuscript better.
I acknowledge the support by the Illinois Survey Science Fellowship from the Center for AstroPhysical Surveys at the National Center for Supercomputing Applications.

This work made use of the Illinois Campus Cluster, a computing resource that is operated by the Illinois Campus Cluster Program (ICCP) in conjunction with the National Center for Supercomputing Applications (NCSA) and which is supported by funds from the University of Illinois at Urbana-Champaign.
This work also used the computational and storage services associated with the Hoffman2 Shared Cluster provided by UCLA Institute for Digital Research and Education's Research Technology Group.

\appendix
\restartappendixnumbering
\section{Signal-to-noise calculation}
\label{appendix_snr_calculation}

\begin{figure*}[b]
\centering
\ifdefined\ApJsubmit
\includegraphics[width=0.9\textwidth, keepaspectratio]{fit_histograms.pdf}
\else
\includegraphics[width=0.9\textwidth, keepaspectratio]{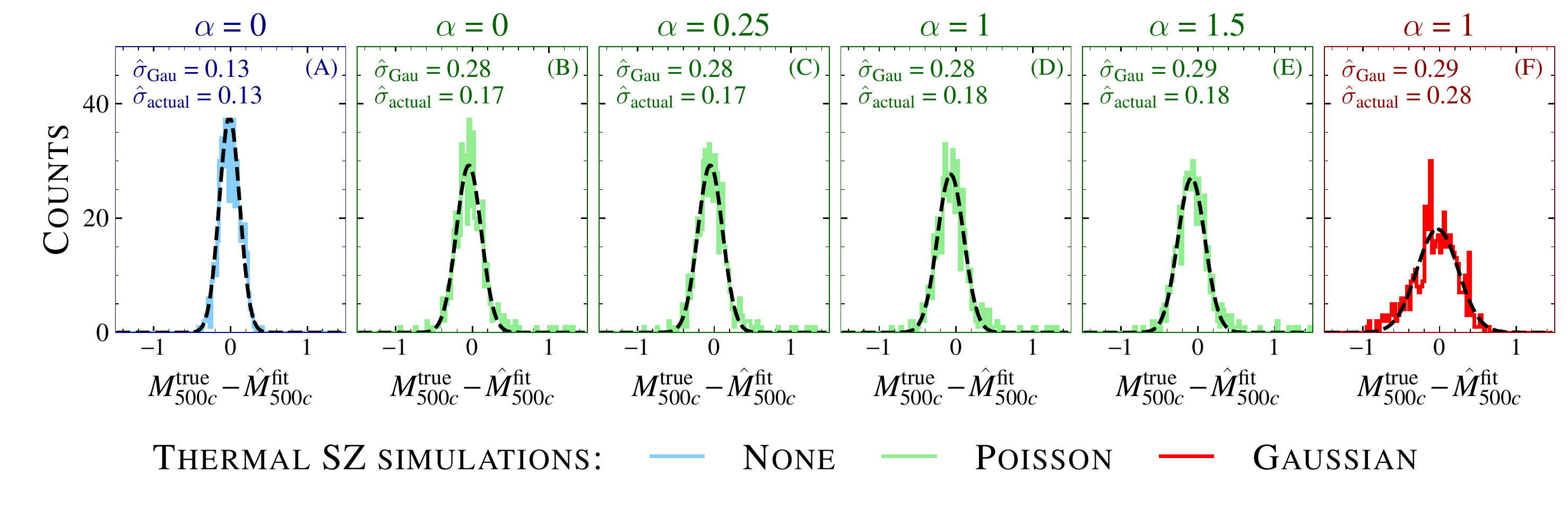}
\fi
\caption{Distribution of the difference between the true and best-fit mass values from \howmanysimulations{} simulations for the three cases of tSZ simulations: blue (no), green (Poisson), and red (Gaussian). 
For blue and red, the distributions are well fit by a Gaussian with $\sigma= \sigma_{\rm Gau}$ but the non-Gaussian tails are clear for green (Poisson case). 
The actual $\sigma_{\rm actual}$ calculated using the $16^{\rm th}$ and the $84^{\rm th}$ percentiles of the distributions is also shown for all cases: $\sigma_{\rm actual} = \sigma_{\rm Gau}$ for blue and red but $\sigma_{\rm actual} < \sigma_{\rm Gau}$ for green.
For the Poisson case, we show the distributions for four different scalings of $\mathbf{\hat{C}}_{\rm tSZ}^{\rm Gaussian}$:  $\gautszcovsuppressionfactor \in [0, 0.25, 1, 1.5]$. 
While the choice of $\gautszcovsuppressionfactor$ modifies the non-Gaussian tails for green panels, it does not affect $\sigma_{\rm actual}$. 
We pick a cluster with $(\mvir, z ) = (0.794\munits, 0.7)$ for this demonstration and use the best-fit distributions from the \cmbhd{} experiment. 
}
\label{fig_fit_histograms}
\end{figure*}

We discuss the $\snr$ calculation for the three cases of tSZ simulations here. 
Fig.~\ref{fig_fit_histograms} shows the distribution of the best-fit values from \howmanysimulations{} simulations for the three kinds of tSZ simulations: no-tSZ in blue (panel A), Poisson-tSZ in green (panels B through E), and Gaussian-tSZ in red (panel F). 
We report the best-fit values in terms of $\mvir$ here. 
The $16^{\rm th}$ and the $84^{\rm th}$ percentiles values of these distributions are used to estimate the $\snr$ in each case. 
We choose this rather than the standard deviations to account for the non-Gaussian tails, which is evident from the figure for the Poisson cases in green. 
The widths calculated this way are $\sigma_{\rm actual}$ and marked in all panels. 
The standard deviations of the distributions, assuming Gaussianity, are $\sigma_{\rm Gau}$. 
As expected, $\sigma_{\rm Gau} = \sigma_{\rm actual}$ for blue (no-tSZ) and red (Gaussian-tSZ) but $\sigma_{\rm Gau} > \sigma_{\rm actual}$ for green (Poisson-tSZ). 
To guide the eye, we also show a simple Gaussian fit with width $\sigma = \sigma_{\rm actual}$ using a black dashed line. 
For this illustration, we choose a cluster with $(\mvir, z ) = (0.794\munits, 0.7)$ as seen by the \cmbhd{} experiment. 
Note from the figure that the $\snr = 1/\sigma_{\rm actual} = 3.6$ for this cluster is below the detection limit for the Gaussian-tSZ case but improves to $\snr = 5.9$ when we switch to the Poisson-tSZ simulations. 

The values of $\gautszcovsuppressionfactor$ marked as titles are the factors applied to scale the tSZ covariance matrix $\mathbf{\hat{C}}_{\rm tSZ}^{\rm Gaussian}$ used in the likelihood calculation (see Eq: \ref{eq_cov_matrix_components}) and estimated using Gaussian simulations of the tSZ signal. 
For the no-tSZ (panel A in blue) and Gaussian-tSZ (panel F in red) cases, we set $\alpha = 0$ and $\alpha = 1$. 
For the Poisson case, since the map variance is dominated by the few pixels with clusters, we test the results by arbitrarily scaling the $\mathbf{\hat{C}}_{\rm tSZ}^{\rm Gaussian}$ using different values of $\alpha \in [0, 0.25, 1, 1.5]$. 
While the choice of  $\alpha$ modifies the non-Gaussian tails, it does not affect $\sigma_{\rm actual}$ as can be inferred from green colored panels (B) through (E). 
This simple scaling does not handle the effect of non-Gaussianities. 
However, given that we do not observe any bias, we assume the impact of the non-Gaussianities to be negligible and ignore them.


\bibliography{tsz}

\end{document}